\newcommand\beq{\begin{eqnarray}}
\newcommand\eeq{\end{eqnarray}}
\def\lsim{\mathrel{\rlap{\lower4pt\hbox{$\sim$}}
    \raise1pt\hbox{$<$}}}                
\def\gsim{\mathrel{\rlap{\lower4pt\hbox{$\sim$}}
    \raise1pt\hbox{$>$}}}

\documentclass[
amsmath,
prd,nofootinbib,floatfix,11pt
]{revtex4}

\allowdisplaybreaks
\usepackage{graphicx}
\usepackage{setspace}

\begin{document}
\renewcommand{\theequation}{\arabic{section}.\arabic{equation}}

\title{\Large%
\baselineskip=21pt
The top squark-mediated annihilation scenario
and direct detection of dark matter in compressed supersymmetry}

\author{Stephen P. Martin}
\affiliation{{\it Department of Physics, Northern Illinois University,
DeKalb IL 60115} and\\
{\it Fermi National Accelerator Laboratory,
P.O. Box 500, Batavia IL 60510}}

\begin{abstract}\normalsize \baselineskip=15pt 
Top squark-mediated annihilation of bino-like neutralinos to top-antitop 
pairs can play the dominant role in obtaining a thermal relic dark matter 
abundance in agreement with observations. In a previous paper, it was 
argued that this can occur naturally in models of compressed 
supersymmetry, which feature a running gluino mass parameter that is 
substantially smaller than the wino mass parameter at the scale of 
apparent gauge coupling unification. Here I study in some more detail the 
parameter space in which this is viable, and compare to other scenarios 
for obtaining the observed dark matter density. I then study the 
possibility of detecting the dark matter directly in future experiments. 
The prospects are consistently very promising for a wide variety of model 
parameters within this scenario.
\end{abstract}


\maketitle

\tableofcontents

\vfill\eject
\baselineskip=15pt

\setcounter{footnote}{1}
\setcounter{page}{2}
\setcounter{figure}{0}
\setcounter{table}{0}

\section{Introduction}
\label{sec:intro}
\setcounter{equation}{0}
\setcounter{footnote}{1}

Supersymmetry provides a candidate for 
the particle dark matter required by cosmological and astrophysical 
observations \cite{DMreviews}, provided that the lightest supersymmetric 
particle (LSP) is a neutralino and $R$-parity is conserved. Supersymmetry 
also contains a solution for the hierarchy problem associated with the 
small ratio of the electroweak breaking scale to the Planck scale and 
other high scales. The simplest version, the Minimal Supersymmetric 
Standard Model (MSSM) \cite{primer}, contains a large undetermined 
parameter space, due mainly to our present ignorance of the mechanism 
behind supersymmetry breaking. In recent years, the parameter space of 
the theory has become increasingly constrained by measurements of and 
limits on flavor violation, searches for Higgs scalar bosons, the density 
of dark matter as inferred from cosmology, and ongoing direct searches 
for superpartners at colliders.

The problem of flavor violation in the MSSM can be solved by assuming 
that the soft supersymmetry-breaking terms are governed by an organizing 
principle that respects flavor. In the most general version, this means 
that at some special input renormalization scale $Q_0$, the 3$\times$3 
squared mass matrices for squarks and sleptons with the same electroweak 
quantum numbers are each (approximately) proportional to the identity matrix:
\beq
{\bf m^2_{\tilde Q}} = m^2_{{\tilde Q},0} {\bf I},\qquad
{\bf m^2_{\tilde u}} = m^2_{{\tilde u},0} {\bf I},\qquad
{\bf m^2_{\tilde d}} = m^2_{{\tilde d},0} {\bf I},\qquad
{\bf m^2_{\tilde L}} = m^2_{{\tilde L},0} {\bf I},\qquad
{\bf m^2_{\tilde e}} = m^2_{{\tilde e},0} {\bf I}.
\eeq
Also, in the idealized limit the sfermion-sfermion-Higgs $3\times 3$ 
matrix couplings in the Lagrangian are proportional to the 
corresponding Yukawa couplings:
\beq
{\bf a_u} = A_{u,0} {\bf y_u},\qquad
{\bf a_d} = A_{d,0} {\bf y_d},\qquad
{\bf a_e} = A_{e,0} {\bf y_e}.
\eeq
These assumptions reduce the number of MSSM parameters
to only 14 beyond those already found in the Standard Model.
The other parameters can be taken to be three
independent gaugino masses $M_1$, $M_2$, $M_3$, the ratio of Higgs
vacuum expectation values $\tan\beta$, and the supersymmetric Higgs mass
parameter $\mu$, taken to be real here.

Most of the well-studied scenarios for the MSSM, including
minimal supergravity (mSUGRA), gauge-mediated supersymmetry breaking,
and anomaly-mediated supersymmetry breaking, are special cases of this
flavor-preserving model framework. For example, as popularly 
applied, mSUGRA makes the further assumptions:
\beq
m_{1/2} &\equiv& M_1 = M_2 = M_3
\label{eq:mSUGRAgauginos}
\\ 
m_0^2 &\equiv& m^2_{{\tilde Q},0} =
m^2_{{\tilde u},0} =
m^2_{{\tilde d},0} =
m^2_{{\tilde L},0} =
m^2_{{\tilde e},0},
\\
A_0 &\equiv&
A_{u,0} = A_{d,0} = A_{e,0}.
\label{eq:mSUGRAA}
\eeq
at a scale $Q_0$ usually taken to be the scale of apparent gauge coupling
unification $M_{\rm GUT} \approx
2 \times 10^{16}$ GeV.
However, it is clear that the parameterization 
(\ref{eq:mSUGRAgauginos})-(\ref{eq:mSUGRAA}) 
may be too simplistic to adequately approximate the real world,
even qualitatively. 

One of the strongest constraints on the MSSM parameter space comes 
from the requirement of electroweak symmetry breaking, given the observed 
value of the top quark mass and the fact that a Higgs scalar boson was 
not discovered at LEP. This requires a balance between the 
supersymmetry-preserving and supersymmetry-violating contributions to the 
Higgs scalar squared mass. The apparent tuning required by this balance 
has become known as the supersymmetric little hierarchy problem.

Another issue is that for many otherwise viable MSSM parameters, the 
predicted thermal relic abundance of the predicted bino-like LSP is too 
large, and the universe would have become overclosed (matter dominated 
too early). Conversely, if the LSP is mostly wino-like or higgsino-like, 
the predicted dark matter density is far too small to agree with the 
results of WMAP and other experiments \cite{WMAP}-\cite{PDG}. The 
exceptional cases in the mSUGRA version of the MSSM usually fall into 
four main categories.

First, there is a ``bulk region" of parameter space, in which the LSP is 
mostly bino-like and pair annihilates efficiently due to the $t$-channel 
and $u$-channel exchange of light sleptons. In much of the mSUGRA 
parameter space, this possibility has been ruled out or is being squeezed 
by the searches for the Higgs boson or other superpartners. 
Second, there is a Higgs resonance region \cite{Drees:1992am}, usually 
found at large $\tan\beta$, in which neutralinos pair annihilate through 
the $s$-channel exchange of the pseudo-scalar Higgs boson $A^0$.
Third, the LSP might 
have a significant higgsino component, allowing efficient neutralino pair 
annihilation and sometimes co-annihilation with the heavier charginos and 
neutralinos, to and through weak bosons \cite{Edsjo:1997bg}. This occurs 
prominently in the ``focus point" region of parameter space 
\cite{focuspoint}, for large $m_0$ in mSUGRA.
Fourth, there is a sfermion co-annihilation region \cite{GriestSeckel}, 
where a sfermion (most often a tau slepton \cite{staucoannihilation} in 
mSUGRA, but possibly a top squark 
\cite{stopcoannihilationone}-\cite{stopcoannihilationfive}) happens to be 
slightly heavier than the LSP. Significant numbers of this sfermion will 
then coexist with the LSP around the time of freeze-out, so 
co-annihilations of the sfermion with itself and the LSP will efficiently 
dilute the superpartners and so the eventual dark matter density.

Recently there have been many studies of dark matter properties that go 
beyond the mSUGRA assumptions, allowing non-universal scalar masses 
\cite{Berezinsky:1995cj}-\cite{Evans:2006sj} or gaugino masses 
\cite{Corsetti:2000yq}-\cite{BBPT} at the input scale. For example, one 
can adjust the wino content of the LSP to be big enough to allow for 
efficient co-annihilations of the LSP with the heavier charginos and 
neutralinos. It is a common theme of these works that the parameter space 
in which the thermal relic density of dark matter comes out in agreement 
with experiment can be significantly enlarged by considering the more 
general boundary conditions.

In ref.~\cite{Martin:2007gf}, I proposed another remedy with distinctive
features, that the LSP is predominantly bino-like
and 
the crucial suppression of the thermal relic density of
dark matter is brought about by the process 
\beq
\tilde N_1 \tilde N_1 \rightarrow t \bar t,
\eeq
mostly mediated by $\tilde t_1$ exchange in the $t$ and $u$ channels. In 
contrast to the other quark and lepton final states, this channel does 
not suffer from $p$-wave suppression, because the top quark mass is 
large. The naturalness of this scenario 
is based on the supposition that the running gluino mass 
parameter $M_3$ is significantly smaller than the bino and wino mass 
parameters $M_1$ and $M_2$ at the scale of apparent gauge unification 
$M_{\rm GUT} \approx 2 \times 10^{16}$ GeV. This causes the spectrum of 
physical superpartner masses to be ``compressed" compared to the usual 
mSUGRA, gauge-mediated, or anomaly-mediated scenarios. If $|M_3| < |M_2|$ 
at $M_{\rm GUT}$, then the amount of fine-tuning required of other 
parameters to obtain electroweak symmetry breaking is substantially 
reduced, as was noted long ago in \cite{KaneKing}. This can occur, for 
example, if the gaugino masses at the apparent GUT scale are governed by 
supersymmetry-breaking $F$-terms in a combination of a singlet and an 
adjoint representation of $SU(5)$. It had previously been suggested in 
models with small $M_3$ that the dark matter density can be explained by 
an enhanced higgsino content of $\tilde N_1$, providing for enhanced 
annihilations via $\tilde N_1 \tilde N_1 \rightarrow W^+ W^-$ and $ZZ$ 
\cite{Bertin:2002sq,Baer:2006dz} or by $s$-channel annihilation mediated 
by the pseudoscalar Higgs $A^0$ near resonance 
\cite{Bertin:2002sq,Belanger:2004ag,Mambrini:2005cp}, by co-annihilations 
with the heavier higgsino-like charginos and neutralinos 
\cite{Belanger:2004ag,Baer:2006dz}, or by $s$-channel annihilations to 
$t\overline t$ through the $Z$ boson 
\cite{Bertin:2002sq,Mambrini:2005cp}.

In compressed supersymmetry, it can be natural for the lighter top 
squark, $\tilde t_1$, to be not much heaver than the LSP and even to be 
the next-to-lightest supersymmetric particle (NLSP), because 
renormalization group contributions to $m_{\tilde t_1}$ from the gluino 
mass are smaller. This is in distinction to mSUGRA and similar models, in 
which $|A_{0}|$ has to be much larger and more finely adjusted in order 
to arrange for $\tilde t_1$ to be not much heavier than the LSP 
\cite{stopcoannihilationtwo,stopcoannihilationthree,%
Edsjo:2003us,stopcoannihilationfive}. Then $\tilde N_1 \tilde t_1$ and 
$\tilde t_1 \tilde t_1$ co-annihilations are usually also very important 
in mSUGRA where this can occur, unlike the compressed supersymmetry case 
discussed here.

Although there is no such thing as an objective measure of fine-tuning, I 
argued in ref.~\cite{Martin:2007gf} that for $|M_3| \sim 0.3 |M_2|$ at 
$M_{\rm GUT}$, this scenario can be considered natural compared to other 
possibilities discussed in the literature, since a relatively large range 
of scalar masses gives a prediction for $\Omega_{\rm DM} h^2$ in the 
range allowed by WMAP and other experiments \cite{WMAP}-\cite{PDG}.

In this paper, I will explore some of the features of the compressed 
supersymmetry scenario in more detail. Section \ref{sec:compressed} 
provides a more detailed look at the allowed parameter space, 
concentrating on the case of combined $SU(5)$ singlet and adjoint 
representation $F$-term supersymmetry breaking. In particular, I will 
explore how compressed symmetry breaking connects to other mechanisms for 
obtaining the observed $\Omega_{\rm DM} h^2$, and the dependence on 
scalar trilinear couplings and $\tan\beta$. In section \ref{sec:direct}, 
I study the prospects in compressed supersymmetry for future direct 
detection of dark matter, which turn out to be quite promising. Section 
\ref{sec:outlook} contains some concluding remarks.

\section{Assumptions and parameters}
\label{sec:general}
\setcounter{equation}{0}
\setcounter{footnote}{1}

The parameter space of the MSSM is highly sensitive to the top-quark mass 
$m_t$, especially because of the way it enters into the prediction for 
the lightest Higgs mass $m_h$. The present combined Tevatron result 
\cite{TeVtopmass} is $m_t = 170.9 \pm 1.1$ (statistical) $\pm 1.5$ 
(systematic). For a fixed lower bound on $m_h$, lower $m_t$ will exclude 
more parameter space. In this paper I will be (perhaps) mildly permissive 
by fixing $m_t$ to be at the 1-sigma upper bound, combining the 
statistical and systematic errors in quadrature:
\beq
m_t &=& 172.7\>{\rm GeV}.
\eeq
In all of the models discussed below, the lightest Higgs boson has very
similar production and decay rates as that of a Standard Model Higgs 
boson, for which the LEP bound \cite{LEPHiggsbounds} is 114.4 GeV. However,
I will again be mildly permissive, enforcing only a bound
\beq
m_h &>& 113.0\>{\rm GeV},
\label{eq:mhconstraint}
\eeq
which takes into account that there remain significant theoretical
uncertainties in the prediction for $m_h$ for any given set of model
parameters.

In the following, the constraint from WMAP and other experiments 
\cite{WMAP}-\cite{PDG} on the thermal relic abundance of dark matter is 
taken to be:
\beq
0.09 < \Omega_{\rm DM} h^2 < 0.13.
\label{eq:DMconstraint}
\eeq
The value of $\Omega_{\rm DM} h^2$ for a given model is obtained by using 
the program {\tt micrOMEGAs 2.0.1} \cite{micrOMEGAs} (checked for 
approximate agreement with DarkSUSY \cite{DarkSUSY}) interfaced to the 
supersymmetry model parameters program {\tt SOFTSUSY 2.0.11} 
\cite{softsusy} (checked for approximate agreement with {\tt SuSpect} 
\cite{suspect} and ISAJET \cite{ISAJET}).

It should also be noted that the MSSM contributions to BR($B \rightarrow 
s \gamma$) reduce it from the Standard Model prediction \cite{SMbsgamma} 
of $(3.29 \pm 0.33) \times 10^{-4}$, and can be significant in the models 
discussed below, leading to an apparent discrepancy with the averaged 
measured value $(3.55 \pm 0.24 \pm 0.10 \pm 0.03) \times 10^{-4}$ for 
$E_\gamma > 1.6$ GeV \cite{Barberio:2007cr}. (This has been discussed in 
more detail in \cite{BBPT}.) However, a small amount of flavor violation 
in the scalar trilinear or squark masses can easily accommodate the 
measurements, without altering the other predictions of the model in any 
substantial way. Therefore, this will not be applied as a constraint.

As noted in the Introduction and ref.~\cite{Martin:2007gf}, the essential
features of compressed supersymmetry can be realized in a simple 
one-parameter extension of the well-known mSUGRA model framework.
If one assumes that the $F$-term VEVs that break supersymmetry transform
as a singlet and an adjoint (${\bf 24}$ dimensional) representation of
$SU(5)$, then the gaugino masses can be parameterized 
by \cite{Ellis:1985jn}-\cite{Anderson:1999ui}:
\beq
M_1 &=& m_{1/2} (1 + C_{24})
\label{eq:c24M1}
\\
M_2 &=& m_{1/2} (1 + 3 C_{24})\\
M_3 &=& m_{1/2} (1 - 2 C_{24}) .
\eeq
applied in this paper at $Q=M_{\rm GUT}$. 
The special case $C_{24} = 0$ yields the mSUGRA (or the minimal 
gauge-mediated) prediction for gaugino masses. For simplicity, I will 
also assume a common scalar squared mass $m_0^2$ and a common scalar 
trilinear coupling parameter $A_0$ at $M_{\rm GUT}$. The parameter $\mu$ 
is assumed to be real and positive, in the convention of \cite{primer}, 
for a phase choice in which $M_3$ is also positive. In the models studied 
here, that tends to give better agreement with the experimental result 
for the anomalous magnetic moment of the muon than the Standard Model, 
although by an amount that is not large compared to the present deviation 
of up to 3$\sigma$ \cite{muonamm,PDG}. Since I am not prepared to claim 
that the Standard Model is ruled out by the anomalous magnetic moment of 
the muon, no constraint is applied. However, negative $\mu$ would give a 
worse agreement than in the Standard Model.

\begin{figure}[!tb]
\includegraphics[width=8.0cm,angle=0]{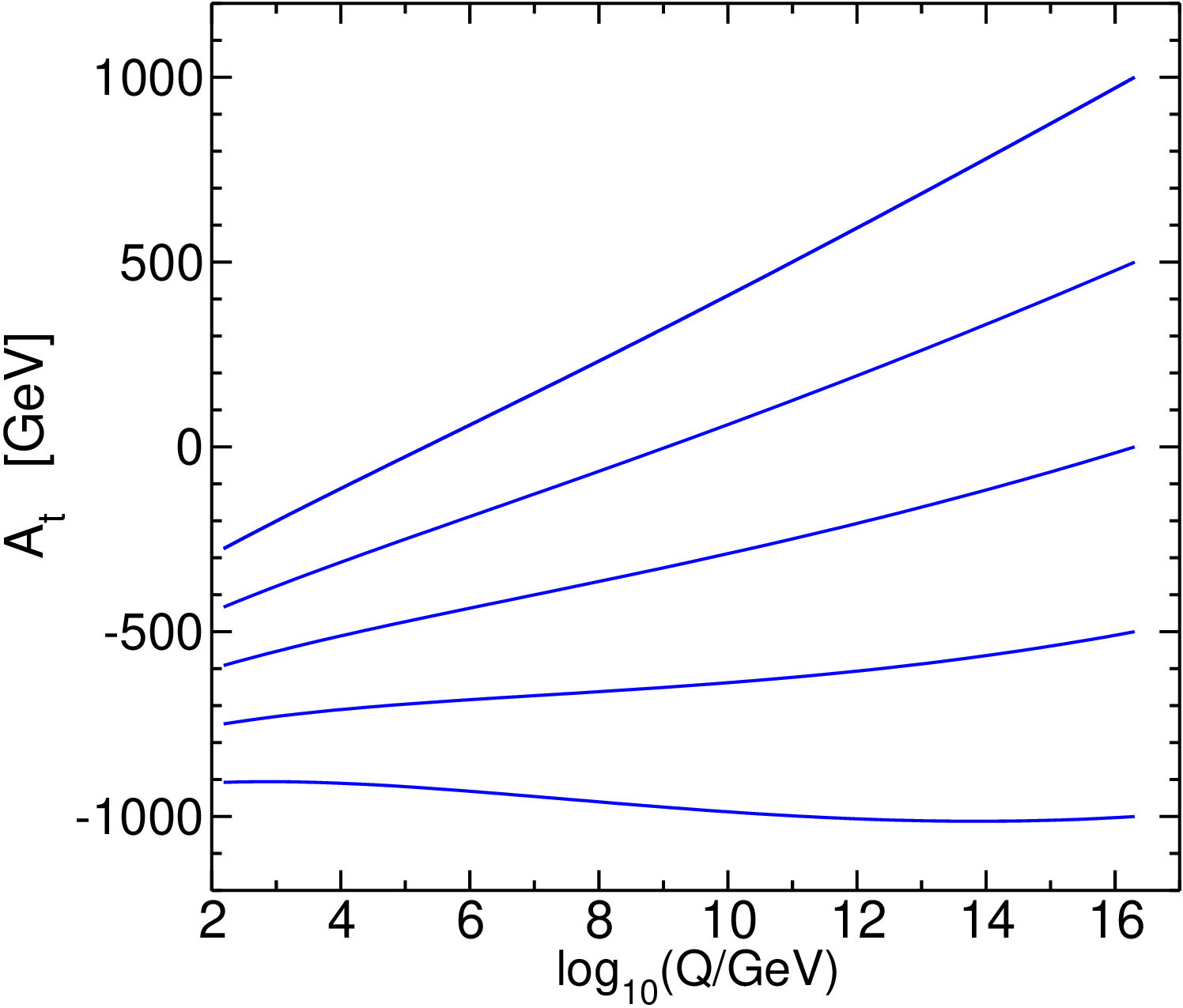}
\caption{\label{fig:Arun} Running of the parameter $A_t \equiv a_t/y_t$
as a function of the renormalization scale $Q$ in compressed 
supersymmetry, for various boundary conditions $A_{0}/M_1 = 0, \pm 1, \pm 
2$ at $M_{\rm GUT} $. The parameters are $C_{24} = 0.21$, $M_1 = 500$ GeV
at $M_{\rm GUT}$,
$\tan\beta = 10$, and $\mu>0$. The renormalization group running
causes $A_t$ to run with a weak focusing behavior
towards negative values at the electroweak scale.} 
\end{figure}

The viability of models considered below requires significant top-squark 
mixing, and a non-zero $A_0$ parameter at $M_{\rm GUT}$. In the 
following, I will assume that $A_0$ is negative, in the conventions of 
\cite{primer}. The motivation for this is that $A_0$ represents the value 
at $M_{\rm GUT}$, while the mSUGRA boundary conditions should more 
properly be applied closer to $M_{\rm Planck}$. If there is strong 
renormalization group running between the Planck scale and the apparent 
GUT scale, it is most likely due to gauge interactions and proportional 
to the gaugino masses, since this maintains the crucial property of 
approximate flavor independence. The corresponding renormalization group 
equations drive the effective parameters $A_{0,u}$, $A_{0,d}$, and 
$A_{0,e}$ towards negative values in the infrared at $M_{\rm GUT}$ where 
they enter as boundary conditions. It has been argued that large 
flavor-independent scalar squared masses and trilinear terms can be 
obtained from an infrared-stable fixed point \cite{IRfp} or an 
ultraviolet-stable fixed point \cite{UVfp} behavior above $M_{\rm GUT}$. 
From this point of view, it is unclear whether the three parameters 
$A_{0,u}$, $A_{0,d}$, and $A_{0,e}$ will actually be unified in the 
absence of true gauge group 
unification, but the coupling of top squarks to the Higgs 
field is of the most direct importance, so I will use a single $A_0$
for simplicity. The 
parameter $A_t = a_t/y_t$ 
continues to run negative below the unification scale. For the models to 
be studied below, $A_t$ does not approach a true fixed point running, but 
has a focusing behavior similar to that well-known in mSUGRA \cite{Atfp}, 
as shown 
in Figure \ref{fig:Arun}. This makes it easy to achieve the required 
level of top-squark mixing for $A_0$ negative and of order $M_1$ at the 
GUT scale, just as expected from \cite{IRfp} or \cite{UVfp} or more 
generally from any kind of strong running above $M_{\rm GUT}$. To achieve 
the right amount of stop mixing with positive $A_0$ at $M_{\rm GUT}$ 
instead would require a large hierarchy of $A_0/M_1$ to overcome the 
tendency of $A_t$ to run negative.

In summary, for the sake of simplicity, I will consider models based on 
the parameters:
\beq
m_{1/2},\qquad C_{24},\qquad m_0,\qquad A_0,\qquad \tan\beta,
\eeq
imposed at $M_{\rm GUT}$, with $\mu>0$ in all cases. I expect that 
similar features will obtain in models with more general boundary 
conditions, with the gaugino mass ratios being the most essential 
qualitative feature.

\section{Compressed supersymmetry}
\label{sec:compressed}
\setcounter{equation}{0}
\setcounter{footnote}{1}

In this section, I present a more detailed look at the compressed MSSM 
parameter space with $C_{24}\not= 0$ in which the thermal dark matter 
density constraint is satisfied, augmenting the discussion in 
ref.~\cite{Martin:2007gf}.

\begin{figure}[!p]
\includegraphics[width=8.07cm,angle=0]{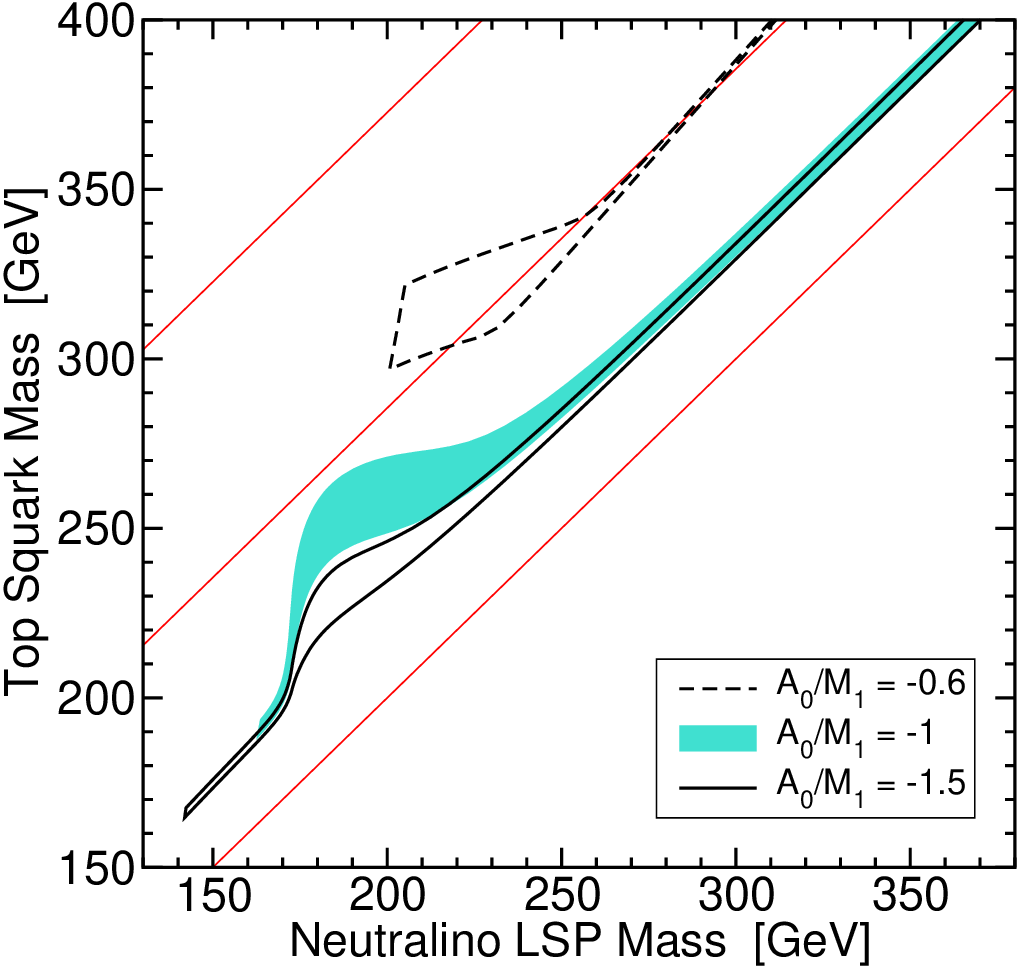}
\hspace{0.05cm}
\includegraphics[width=8.07cm,angle=0]{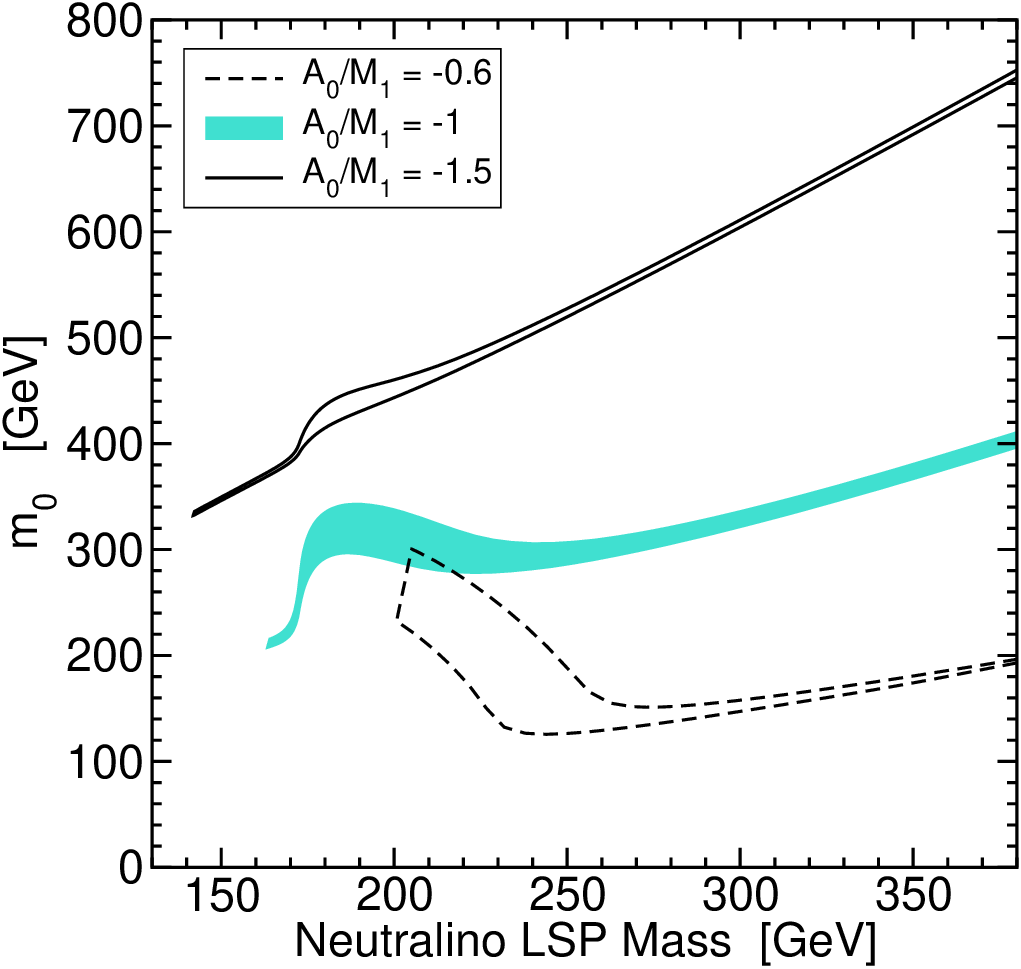}
\caption{\label{fig:exampleA} The regions that satisfy the constraint 
$0.09 < \Omega_{\rm DM} h^2 < 0.13$ are shown for $C_{24} = 0.21$, with 
$\tan\beta = 10$, $\mu>0$ and varying $M_1$ and $m_0$. The regions 
enclosed by dashed lines, shaded, and enclosed by solid lines correspond 
respectively to $A_0/M_1 = -0.6$, $-1$, and $-1.5$ at the unification 
scale. The left panel shows $m_{\tilde t_1}$ as a function of the 
neutralino LSP mass $m_{\tilde N_1}$. The lowest thin (red) line 
corresponds to $m_{\tilde t_1} = m_{\tilde N_1}$, below which the stop 
would be 
the LSP. The middle and upper thin (red) lines bound the regions in which 
respectively $\tilde t_1 \rightarrow W b \tilde N_1$ and $\tilde t_1 
\rightarrow t \tilde N_1$ become kinematically allowed. The right panel 
shows the input parameter $m_0$ as a function of the LSP mass for the 
same regions.}
\vspace{0.9cm}

\includegraphics[width=8.07cm,angle=0]{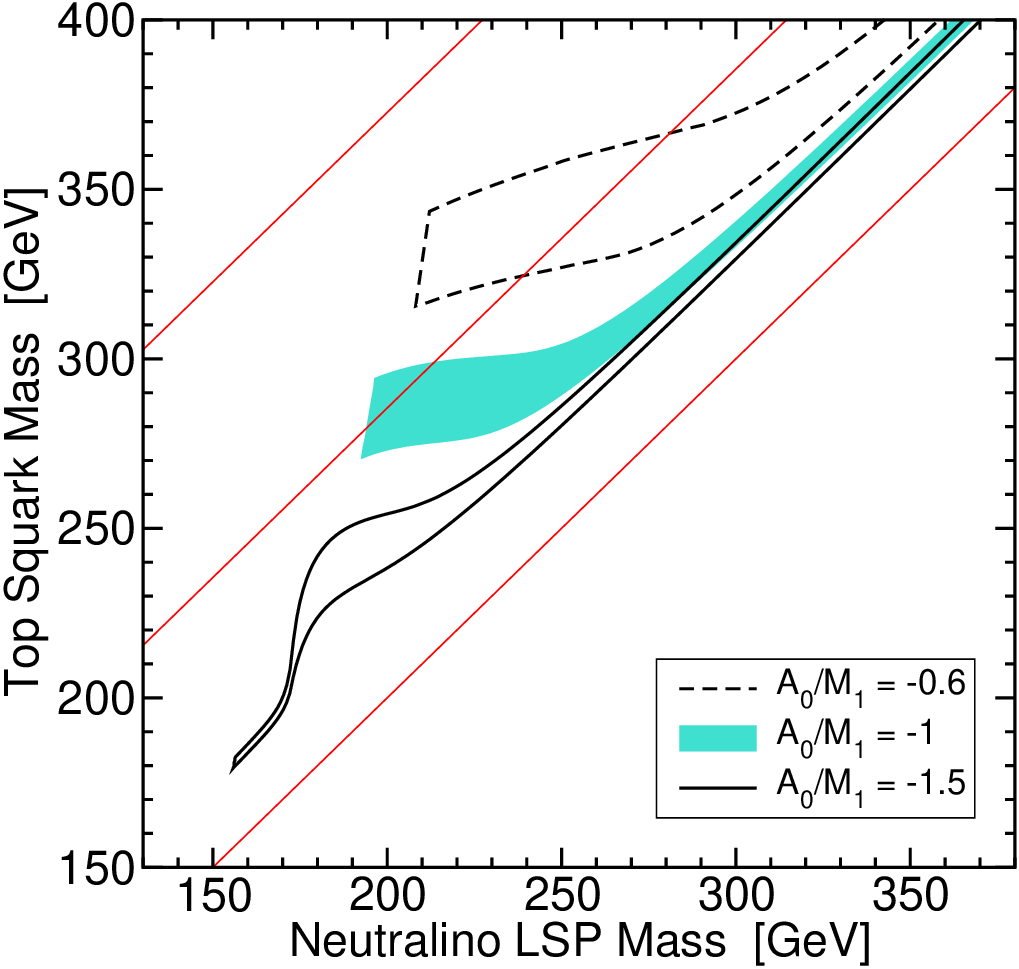}
\hspace{0.05cm}
\includegraphics[width=8.07cm,angle=0]{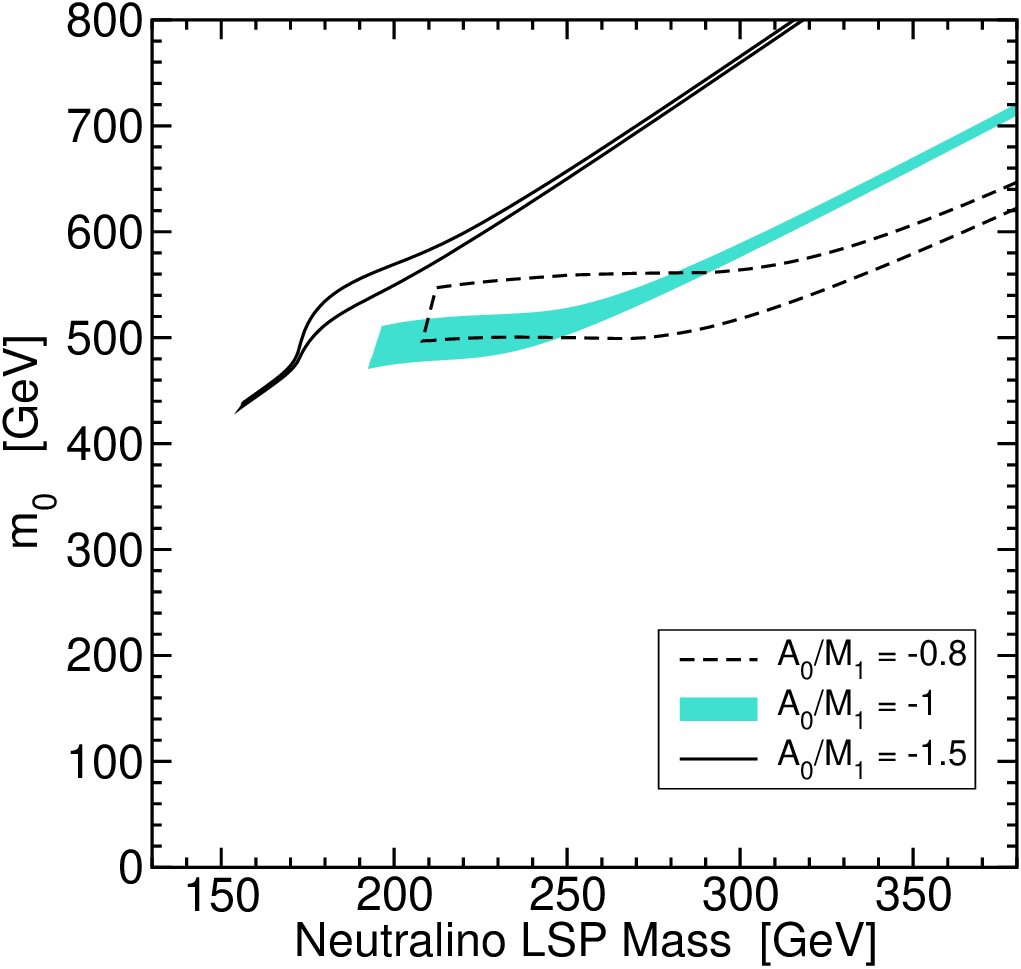}
\caption{\label{fig:exampleB} As in Figure \ref{fig:exampleA}, but with
$C_{24} = 0.24$, and now the regions
enclosed by dashed lines, shaded, and enclosed by solid lines correspond
respectively to
$A_0/M_1 = -0.8$, $-1$, and $-1.5$ at the unification scale.} 
\end{figure}

Figure \ref{fig:exampleA} shows the regions allowed by the constraints on 
$m_h$ and $\Omega_{\rm DM} h^2$ in some typical two-parameter model 
spaces. Here I have chosen $C_{24} = 0.21$, which implies approximately 
$M_1 : M_2 : M_3 :: 2.04 : 2.81 : 1$ at the apparent unification scale 
$M_{\rm GUT}$. The other fixed parameters are $\tan\beta = 10$, $\mu>0$, 
and three values $A_0/M_1 = -0.6, -1.0,$ and $-1.5$
at $M_{\rm GUT}$. The parameters $M_1$ 
and $m_0$ are free. The bulge regions are where $\tilde N_1 \tilde N_1 
\rightarrow t \bar t$ makes the dominant contribution to reducing the 
dark matter density. The left panel shows the allowed values of the 
lighter top squark mass $\tilde t_1$, subject to the dark 
matter constraint of 
eq.~(\ref{eq:DMconstraint}). The Higgs mass bound 
eq.~(\ref{eq:mhconstraint}) cuts off the allowed region on the left. 
This constraint becomes more important for smaller values of $-A_0/M_1$, 
cutting away much of the bulge region for the $-A_0/M_1 = 0.6$ case. For 
the thinner regions with $m_{\tilde N_1}$ above and below the bulge 
regions, stop co-annihilations (for $-A_0/M_1 = 1.0$ and $1.5$) or stau 
co-annihilations (for $-A_0/M_1 = 0.6$) rather than annihilations to top 
quarks are the crucial factor in limiting the dark matter density.

In the right panel of Figure \ref{fig:exampleA}, I show the values of the 
input parameter $m_0$ required for the same models. Here we note the 
trend that for larger $-A_0/M_1$, the allowed regions require $m_0$ to be 
larger and in a narrower range. This finer required adjustment may be 
taken to imply that much larger values of $-A_0/M_1$ are not as likely. 
Conversely, for smaller values of $-A_0/M_1$, the region that would be 
otherwise allowed is eliminated by the Higgs mass constraint. (Recall 
that I have already been mildly permissive in this regard.) One can 
conclude that moderately negative values of $-A_0/M_1$ at the GUT scale 
are the most likely realizations of the compressed supersymmetry 
scenario.

Figure \ref{fig:exampleB} shows similar plots, but this time for the case 
of $C_{24} = 0.24$, which implies the somewhat more severe hierarchy $M_1 
: M_2 : M_3 :: 2.38 : 3.31 : 1$ at the GUT scale. This time, because the 
ratio of squark to neutralino masses is relatively even smaller, the 
Higgs mass bound is correspondingly more stringent, and now cuts off a 
significant part of the $-A_0/M_1 = 1.0$ region. The $-A_0/M_1 = 0.6$ 
bulge region does not survive at all, and so has been replaced by 
$-A_0/M_1 = 0.8$. In the right panel, one sees that the values of $m_0$ 
that are required are larger than for the $C_{24} = 0.21$ case, but still 
only of order $500$ GeV in the stop-mediated bulge region.

\begin{figure}[!tp]
\includegraphics[width=8.07cm,angle=0]{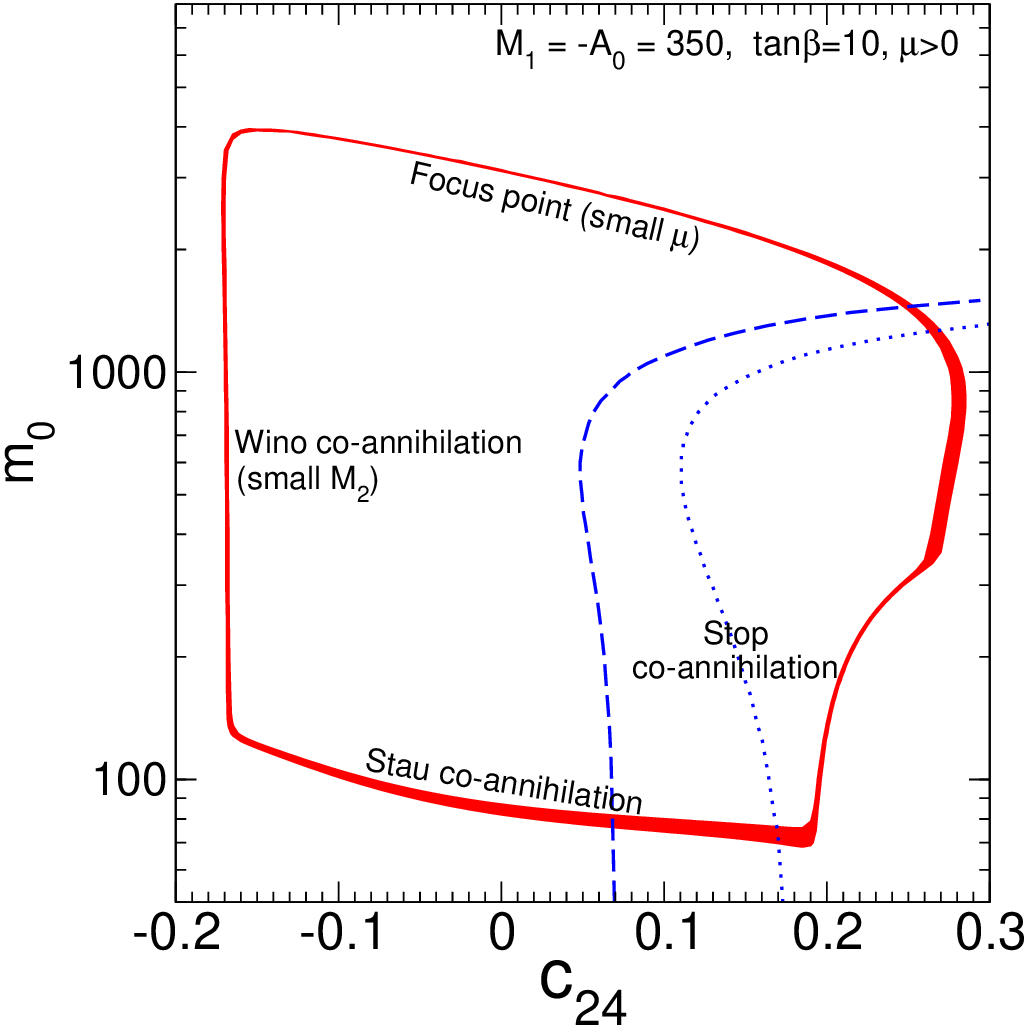}
\includegraphics[width=8.07cm,angle=0]{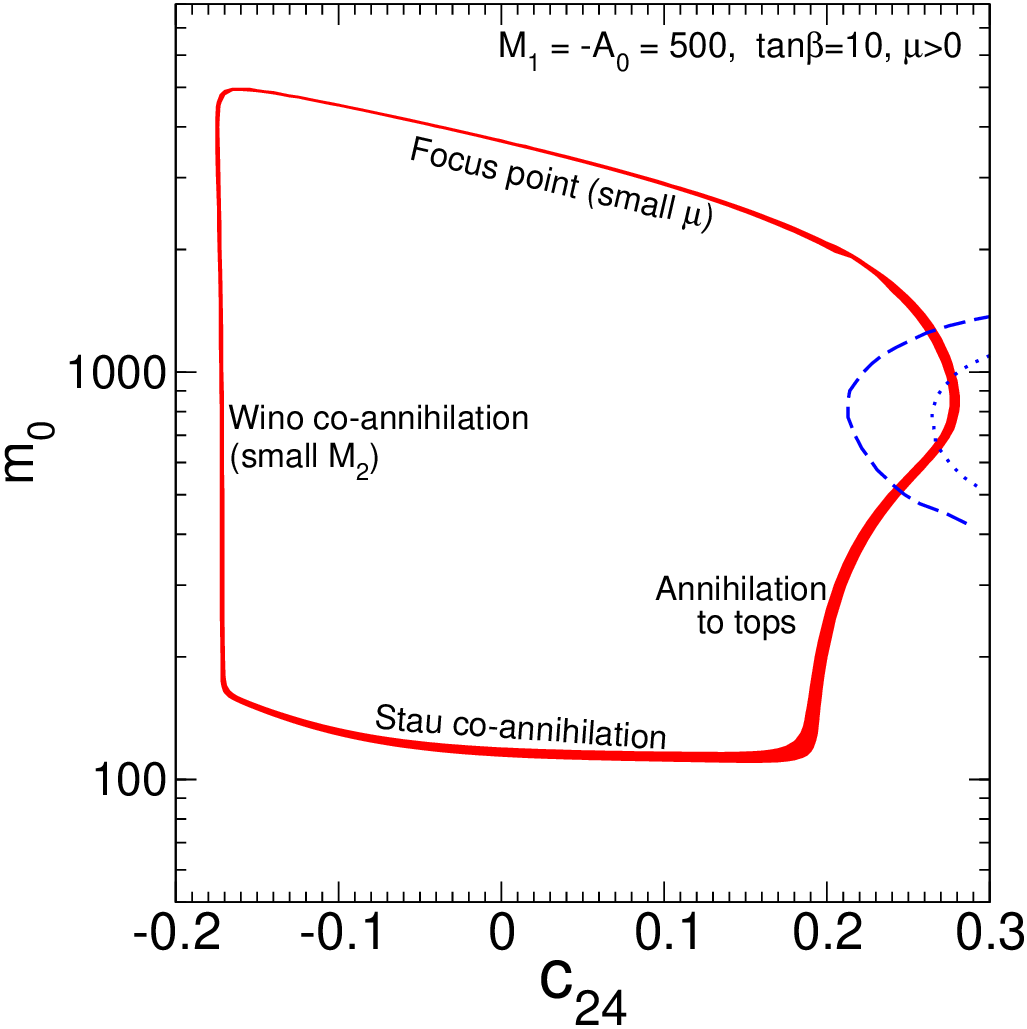}
\vspace{0.3cm}
\includegraphics[width=8.07cm,angle=0]{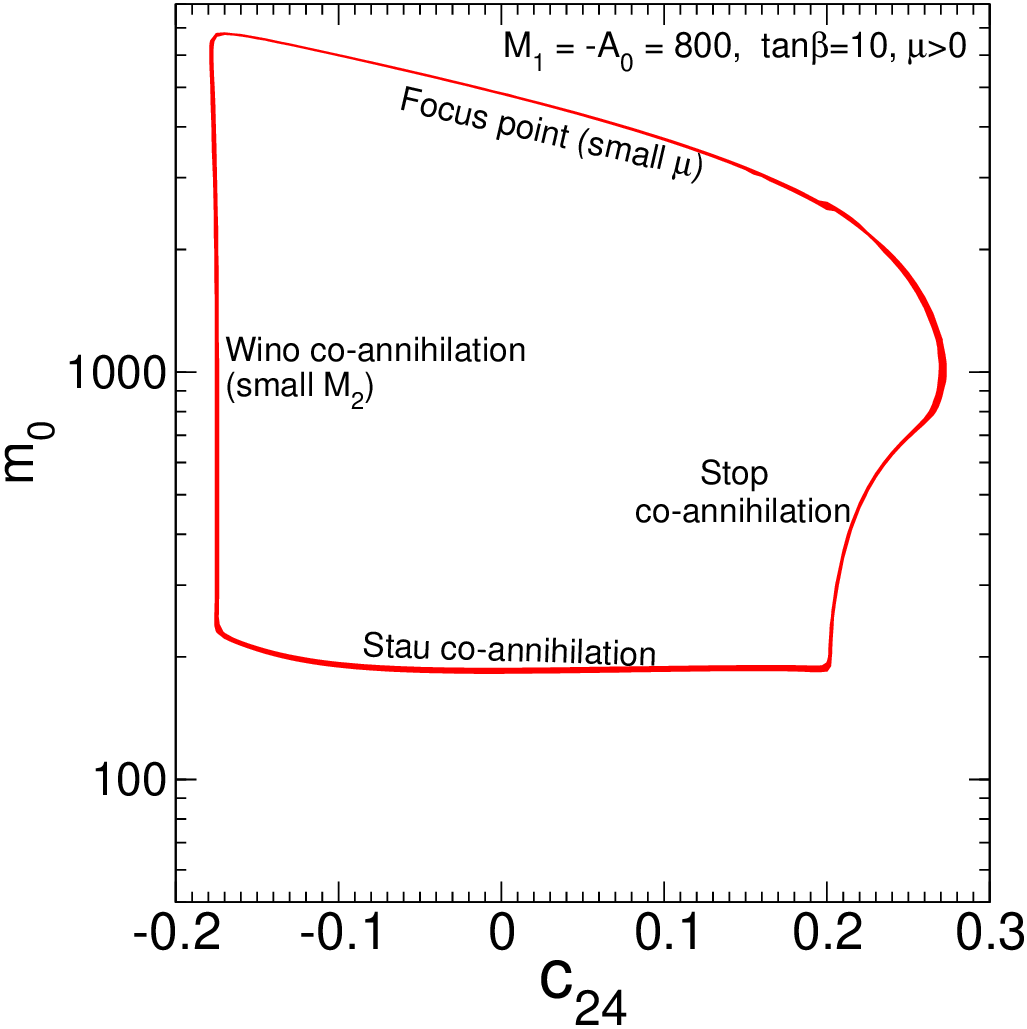}
\caption{\label{fig:bigpicture} The shaded (red) regions of parameter space
shown are allowed by the thermal relic dark matter density constraint
$0.09 < \Omega_{\rm DM} h^2 < 0.13$. 
The three panels show results for $M_1 = -A_0 =
350$, 500, and 800 GeV, with $\tan\beta=10$ and $\mu>0$
in all cases, and with varying common scalar masses $m_0$ and
gaugino non-universality parameter $C_{24}$. In the $M_1 = 350$ and
$M_1 = 500$ GeV
panels, all points to the right of the dashed (blue) 
line have a computed value of $m_h < 113$ GeV, and all points to the 
right of the
dotted (blue) line have $m_h < 112$ GeV.
The $M_1 = 500$ GeV case has a region in the lower right where
$\tilde N_1 \tilde N_1 \rightarrow t \overline t$ dominates
the determination of $\Omega_{\rm DM} h^2$ without significant
assistance from co-annihilations, as suggested in ref.~\cite{Martin:2007gf}.}
\end{figure}

In order to put the scenario illustrated above into a more general 
context, consider the dark matter allowed regions as $C_{24}$ and $m_0$ 
are allowed to vary. Figure \ref{fig:bigpicture} shows the allowed 
regions as shaded (red) for three fixed values of $M_1 = -A_0 = 350$, 
500, and 800 GeV, with $\tan\beta = 10$ and $\mu>0$. The allowed region 
in each case consists of a thin strip, topologically forming a ring, with 
long sections where the qualitative features leading to the correct dark 
matter density are relatively constant and identifiable.

For $C_{24}=0$ (the mSUGRA case), there are two solutions, one with very 
large $m_0$ of order several TeV, and the other with small $m_0$. The 
large $m_0$ solution is the focus point scenario \cite{focuspoint}, in which $\mu$ is small 
and the LSP has a significant Higgsino content. As $m_0$ is increased beyond 
this large value, $\mu$ becomes even smaller and there is soon no 
solution consistent with LEP bounds or electroweak symmetry breaking. The 
solution for $C_{24} = 0$ with small $m_0$ is a stau 
co-annihilation region, in which the mass difference between the LSP and 
the stau is sufficiently small that they exist together in thermal 
equilibrium in the early universe, allowing efficient co-annihilations. 
Both the focus point and stau co-annihilation regions exist for a 
significant range of both positive and negative $C_{24}$.

The nearly vertical left side of the allowed region in each case, near 
$C_{24} = -0.17$ to $-0.18$, corresponds to $M_2/M_1 
\sim 0.6$ near $M_{\rm GUT}$. This leads to an LSP with a large wino 
content, which in the allowed region is tuned to give just the right dark 
matter density. In this region, the co-annihilations of $\tilde N_1 
\tilde C_1$ and $\tilde C_1 \tilde C_1$ and $\tilde N_1 \tilde N_1$ and 
$\tilde N_2 \tilde C_1$ are all important.

In the lower right part of the $M_1 = 350$ GeV
panel, there is a thin stop co-annihilation region where the stop is the 
NLSP. However, the Higgs mass bound, shown as a dashed (blue) line
for $m_h > 113$ GeV and a dotted (blue) line for the more optimistic
constraint $m_h 
> 112$ GeV, 
eliminates this region (and the part of the stau co-annihilation region 
with $C_{24} > 0.07$) for the smaller $M_1$ case.

For the case of intermediate gaugino masses, illustrated in Figure 
\ref{fig:bigpicture} by $M_1= 500$ GeV, the stop co-annihilation region 
is replaced by a fatter region where $\tilde N_1 \tilde N_1 \rightarrow t 
\bar t$ mediated by $\tilde t_1$ is the dominant mechanism for dark 
matter suppression. This scenario is the main object of interest here; it 
is the one that was argued for in \cite{Martin:2007gf}, and appears as 
the bulge regions of Figures \ref{fig:exampleA} and \ref{fig:exampleB}. 
It is important to note that in this region, co-annihilations play only a 
very small role. Only the upper right corner of this region, where it 
would join with the focus point region, is eliminated by the Higgs mass 
constraint.

For sufficiently large $M_1$, the LSP becomes too heavy for
$\tilde N_1 \tilde N_1 \rightarrow t
\bar t$ to dominate, and the thin stop co-annihilation region
reappears, as illustrated in
the lower right part of the $M_1 = 800$ GeV panel of 
\ref{fig:bigpicture}. In this case, the $m_h$ constraint does not
have any impact, but the price to be paid for this is that the
tuning required to obtain correct electroweak symmetry breaking
and the tuning required to obtain the right amount of dark matter
are both worsened.

It should be remarked that in the focus point region with $m_{\tilde N_1} 
> m_t$, it is often the case that the process $\tilde N_1 \tilde N_1 
\rightarrow t \bar t$ is 
also important in regulating the dark matter density, although for a 
quite different reason; there the top squark is irrelevant, and the 
$s$-channel $Z$ and Higgs exchange diagrams (due to the higgsino content 
of the LSP) are the important ones.

In each of the three panels in Figure \ref{fig:bigpicture}, $\Omega_{\rm 
DM} h^2$ is predicted to be too large in the unshaded interior region, if 
the dark matter is due to thermal relics.

\begin{figure}[!tp]
\includegraphics[width=8.1cm,angle=0]{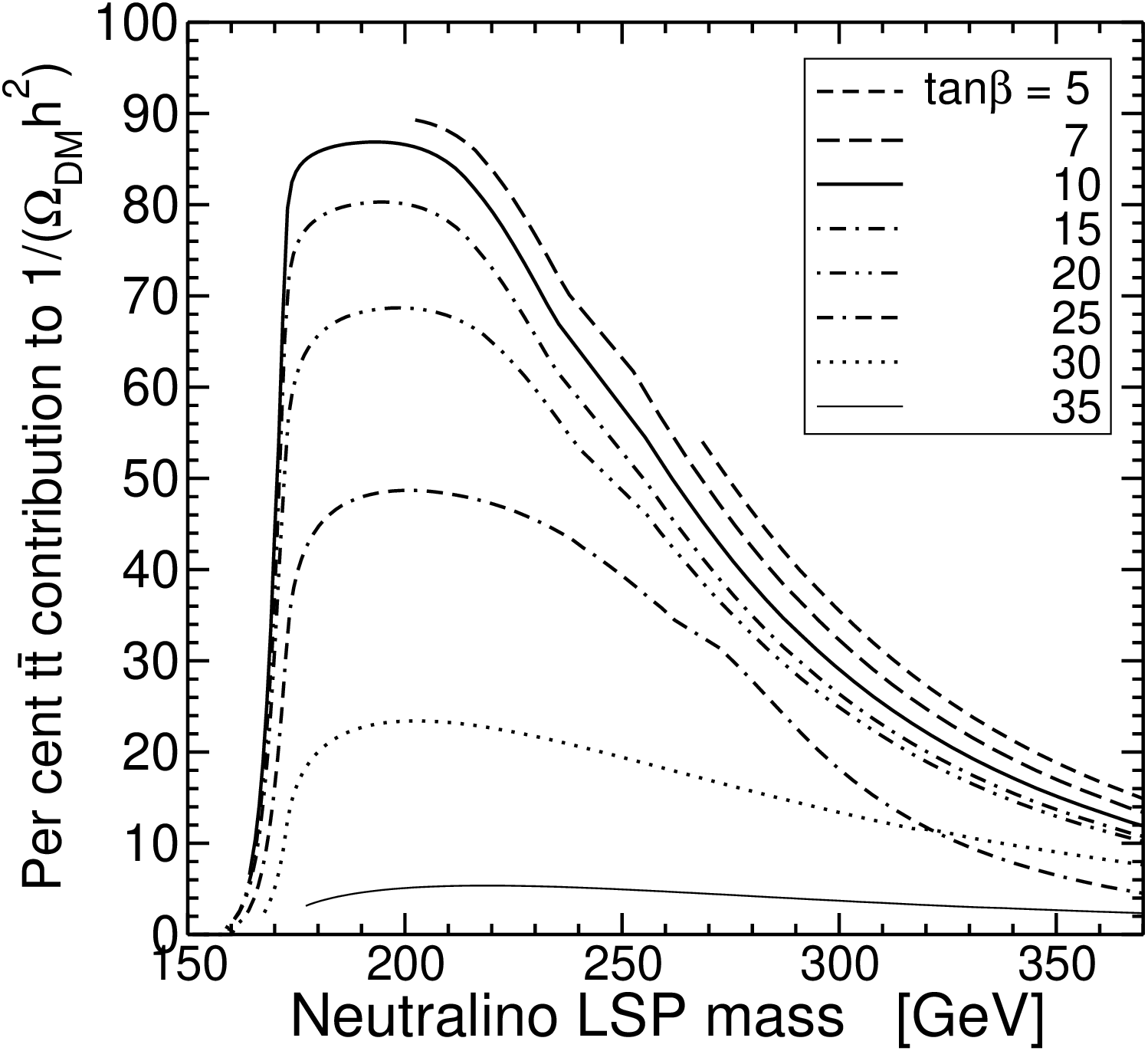}
\caption{\label{fig:tanbeta}The relative contribution of the process
$\tilde N_1 \tilde N_1 \rightarrow t \bar t$ to 
$1/(\Omega_{\rm DM} h^2)$, in per cent, for various values of
$\tan\beta$, as a function of $m_{\tilde N_1}$. In all cases, $C_{24} = 
0.21$, $A_0 = -M_1$ at $Q = M_{\rm GUT}$, $\mu>0$, $M_1$ is varying,
and $m_0$ is adjusted to obtain $\Omega_{\rm DM} h^2 = 0.11$.}
\end{figure}

For larger $\tan\beta$, the stau co-annihilation region requires larger 
scalar masses
$m_0$, since the lighter stau mass is reduced by the effects of 
the tau Yukawa coupling. This tends to squeeze out the region in which a 
light stop can play an important role, as $\tilde N_1 \tilde N_1
\rightarrow \tau^+\tau^-$ mediated by staus becomes more important. 
Furthermore, the annihilation of 
LSPs through the pseudo-scalar Higgs $A^0$ eventually opens up for larger 
$\tan\beta$, so that the relative importance of $\tilde N_1 \tilde N_1 
\rightarrow t \bar t$ is further reduced in the dark matter allowed 
region. Therefore, one expects that for sufficiently large $\tan\beta$, 
the scenario in which annihilation to top quarks plays the most important 
role for small or moderate $m_0$ will disappear. This is illustrated in 
Figure \ref{fig:tanbeta}, where $C_{24} = 0.21$ and $A_0/M_1 = -1$ are 
held fixed, $\mu>0$, and for each value of $M_1$, $m_0$ is adjusted to 
the (lower) value that gives rise to $\Omega_{\rm DM} h^2 = 0.11$. For 
various values of $\tan\beta$, the figure shows the relative contribution 
of the process $\tilde N_1 \tilde N_1 \rightarrow t \bar t$ to 
$1/(\Omega_{\rm DM} h^2)$, in per cent, as a function of $m_{\tilde 
N_1}$.
For $\tan\beta \lsim 10$, the Higgs mass constraint
eliminates viable models unless $m_{\tilde N_1}$ is sufficiently large.
For $\tan\beta \lsim 5$, the LSP is forced to be so heavy that the
annihilation to top quarks is no longer the dominant factor.
For $\tan\beta \gsim 25$, the annihilation to top quarks is again
not the dominant process, and is nearly negligible for $\tan\beta \gsim 
35$.
It is for the intermediate ranges, roughly 
\beq
&&5 \lsim \tan\beta \lsim 25
\\
&&m_t < m_{\tilde N_1} \lsim m_t + 100\>{\rm GeV}
\\
&&m_{\tilde N_1} + 25\>{\rm GeV} \lsim m_{\tilde t_1} 
\lsim m_{\tilde N_1} + 150\>{\rm GeV}
\eeq
that the stop-mediated annihilation of LSPs can play the most important 
role.

\section{Direct detection of dark matter neutralinos}
\label{sec:direct}
\setcounter{equation}{0}
\setcounter{footnote}{1}

Dark matter neutralinos can in principle be detected by their weak 
interaction elastic scattering with ordinary matter nuclei in 
low-background laboratory detectors. The 
spin-independent part of the neutralino-nucleon elastic cross-section 
adds coherently for heavy nuclei, and so gives the best prospects for 
successful detection. In most cases, the $\tilde N_1 p \rightarrow \tilde 
N_1 p$ and $\tilde N_1 n \rightarrow \tilde N_1 n$ cross-sections are 
comparable, and the difference is not large compared to theoretical 
uncertainties associated with the structure of the proton and neutron. It 
has therefore become conventional to describe search results and 
projections in terms of the spin-independent scattering cross-section for 
the dark matter particles on the proton.

\begin{figure}[!p]
\includegraphics[width=8.1cm,angle=0]{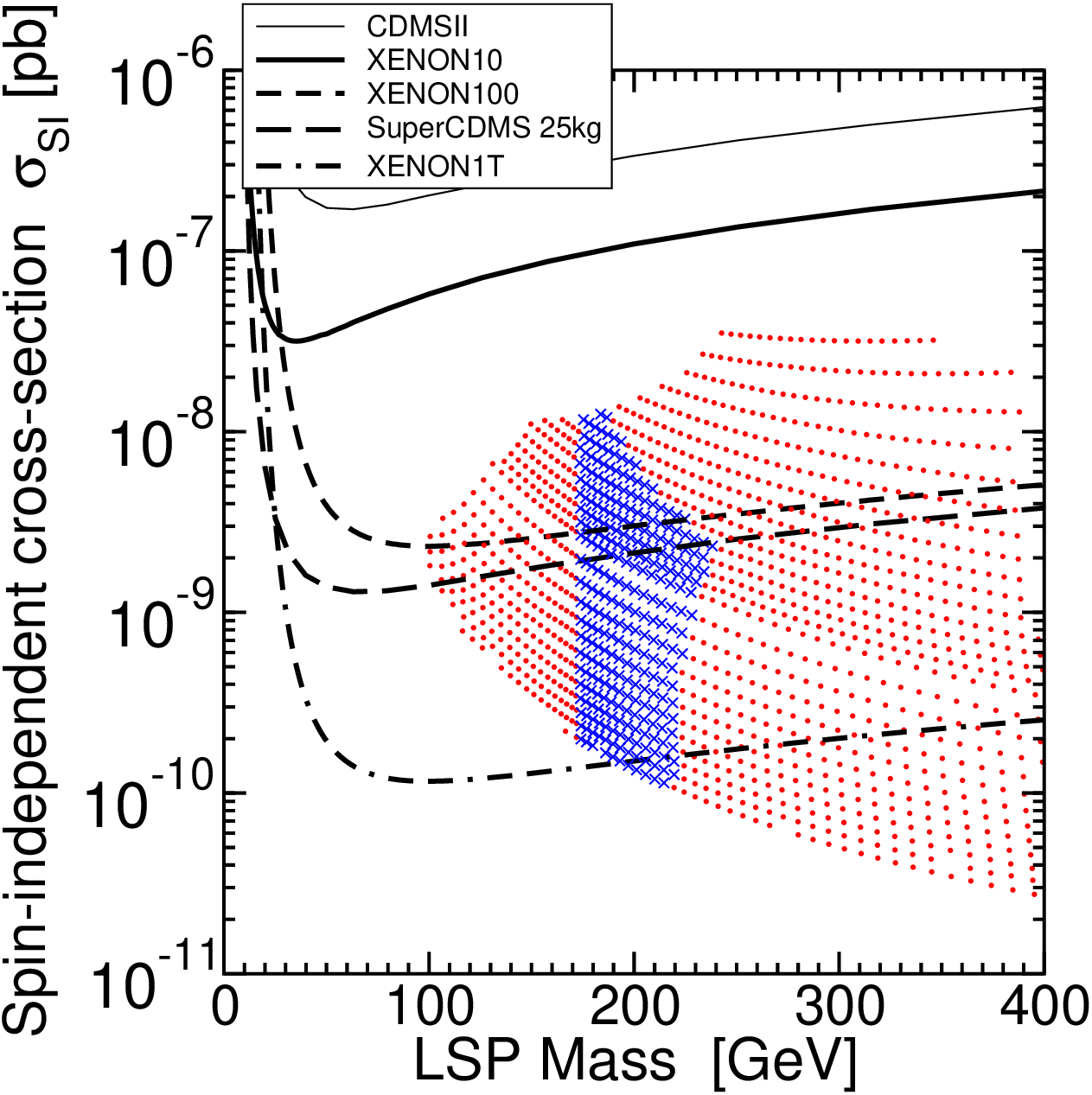}
\includegraphics[width=8.1cm,angle=0]{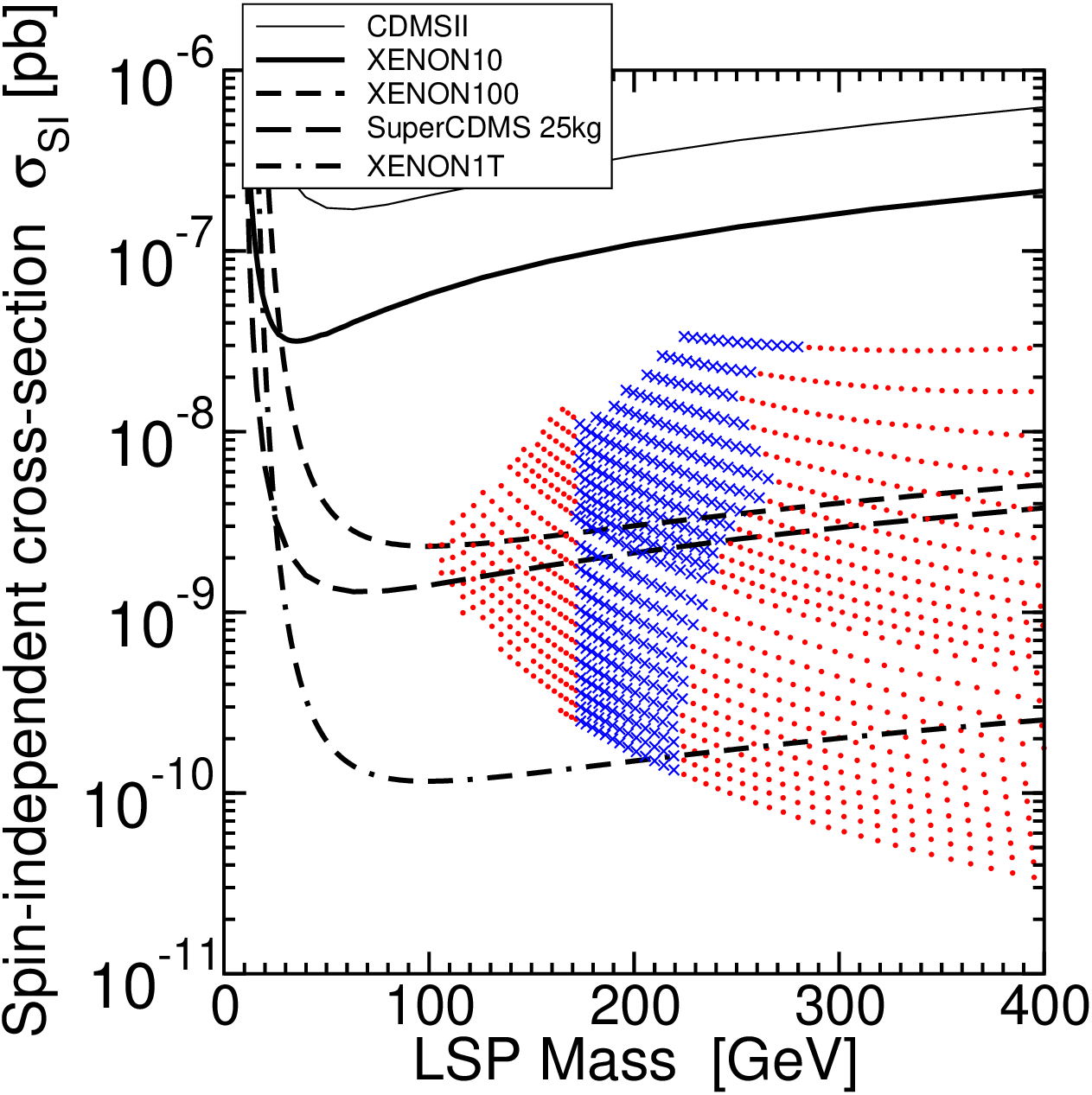}

\vspace{0.4cm}

\includegraphics[width=8.1cm,angle=0]{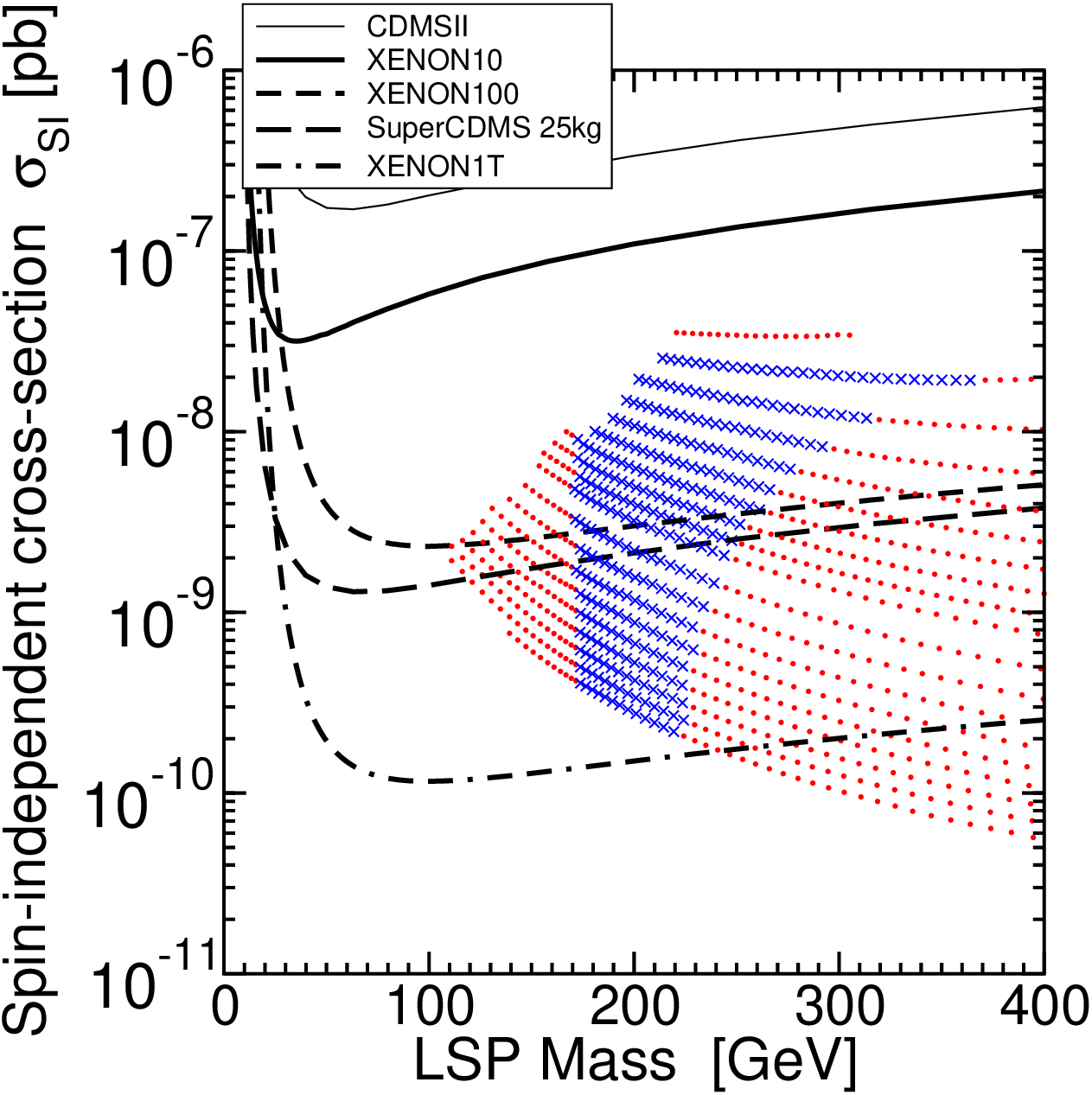}
\includegraphics[width=8.1cm,angle=0]{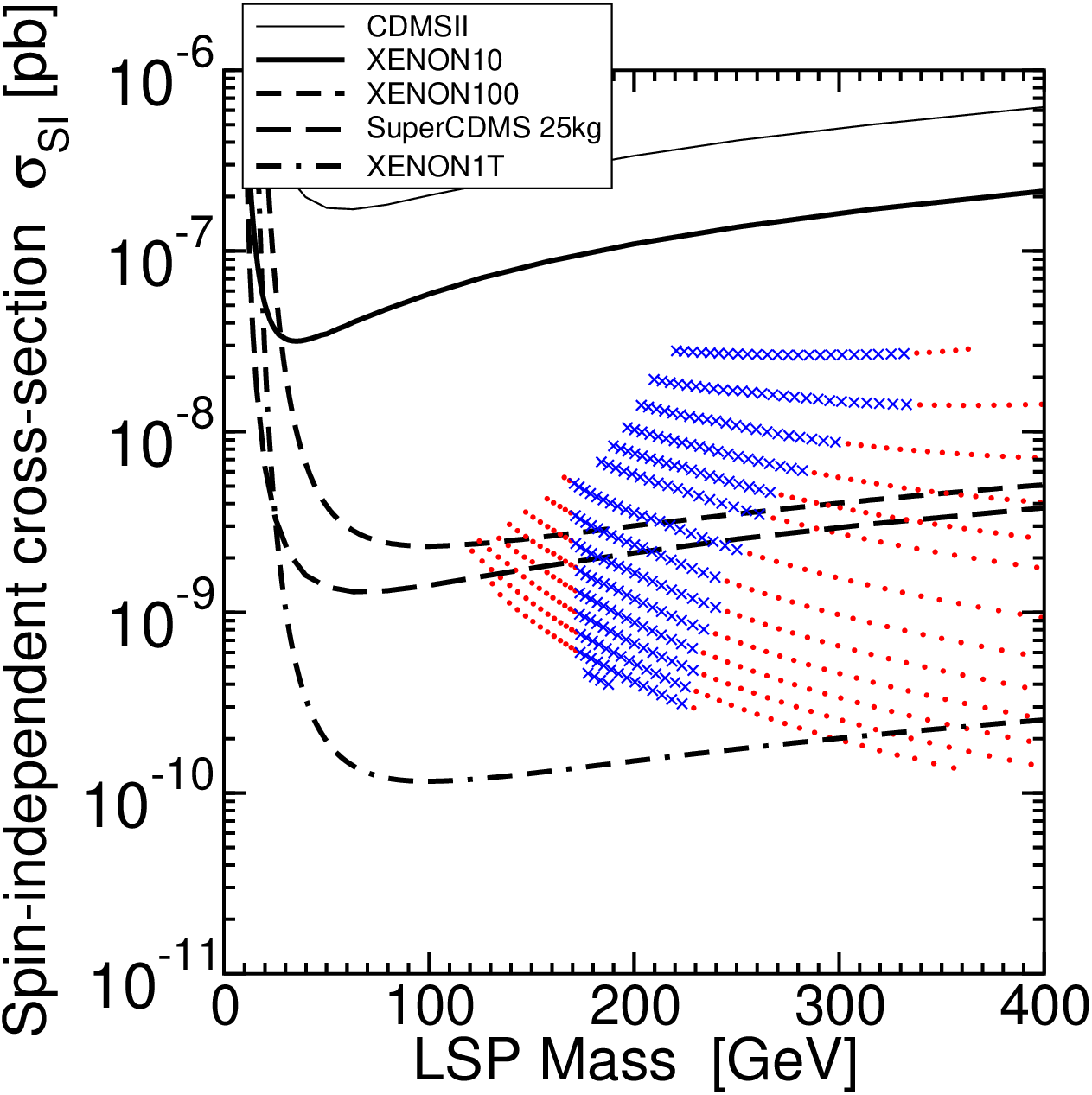}
\caption{\label{fig:DD}Spin-independent
proton-LSP cross-sections in compressed supersymmetry models,
compared to existing limits (solid lines) and some future projected 
reaches (broken lines)
from direct detection experiments.
Models in which $\tilde N_1 \tilde N_1 \rightarrow t \overline t $
contributes at least 50 per cent to $1/(\Omega_{\rm DM} h^2)$ are
denoted by (blue) X's, and other models satisfying the WMAP 
constraints but with $\tilde N_1 \tilde N_1 \rightarrow t \overline t $
contributing less than 50 per cent to $1/(\Omega_{\rm DM} h^2)$
are denoted by (red) dots. The four panels correspond to $C_{24} = 0.19$
(upper left), $0.21$ (upper right), $0.24$ (lower left), and $0.28$
(lower right). 
For each model, $\tan\beta = 10$, $\mu>0$,
and $m_0$ is adjusted to the (lower) value such 
that $\Omega_{\rm DM}h^2 = 0.11$.
The highest line of models in each of the 
$C_{24} = (0.19, 0.21, 0.24, 
0.28)$ panels
corresponds to 
$-A_0/M_1 = (0.1, 0.4, 0.7, 1.1)$ 
respectively. Each lower line of models has
$-A_0/M_1$ increased by $0.1$ until $-A_0/M_1 = 1.6$, 
and then $-A_0/M_1$
increases by $0.2$ for each lower line of models. The lowest line of
models for $C_{24} = (0.19, 
0.21, 0.24,
0.28)$
has $-A_0/M_1 = (4.0, 3.8, 3.4, 3.4)$, respectively. 
These upper bounds on $-A_0/M_1$, which 
set the
lower limits on the cross-section, follow from the requirement of a 
consistent model solution with a stable vacuum and
$\Omega_{\rm DM} h^2 = 0.11$. 
The lower bounds on
$-A_0/M_1$ in each panel come from the requirement $m_h > 113$ GeV.} 
\end{figure}

Figure \ref{fig:DD} shows lines corresponding to present and future 
projected limits from some dark matter direct detection experiments, 
assuming a standard local density of $\rho_{\rm LSP} \approx 0.3$ 
GeV/cm$^3$. The solid lines are the latest results from CDMSII 
\cite{CDMSIIlimit} and XENON10 \cite{XENONten}. Projections for two 
representative next-generation experiments, the XENON100 \cite{XENON} and 
SuperCDMS 25kg at SNOLAB \cite{SuperCDMS}, appear as dashed lines. The 
lowest line is the projection for a ton-scale Xenon experiment, XENON1T 
\cite{XENON}. The data for these lines came from the compilation 
and comparison data archive at \cite{DMtools}.

Also shown in Figure \ref{fig:DD} are the results for the spin-independent 
LSP-proton cross-section in a variety of compressed supersymmetry models. 
The spin-independent cross-section
\beq
\sigma_{SI} (\tilde N_1 p 
),
\eeq
has been computed using an implementation of the formulas of Drees and 
Nojiri \cite{Drees:1993bu}. It should be noted that the prediction is
subject to significant uncertainties
that are presently unavoidable, due in part to the lack of accuracy 
for the values of the quark-antiquark matrix elements for the proton,
\beq
f^p_{T_q} = \langle p |m_q \bar q q | p \rangle/m_p
\qquad\qquad (q = u,d,s).
\eeq
In the following, I use in particular:
\beq
f^p_{T_u} = 0.023,\qquad f^p_{T_d} = 0.034, \qquad f^p_{T_s} = 0.14.
\eeq
Other values used in the literature can give results for $\sigma_{SI}$ 
that are higher or lower by a factor of up to 2 or 3; see for example 
\cite{DDuncertainties}.

The most important contribution to the spin-independent cross-section in 
the models examined below always comes from $h^0$ (the lighter CP-even 
neutral Higgs boson) exchange, with $H^0$ exchange playing a lesser role 
and other contributions usually still less. The cross-section is enhanced 
by the bino-higgsino mixing that is generally a feature of compressed 
supersymmetry, due to $|\mu|$ not being too large. It has already been 
noted, for example in \cite{Baer:2006dz}, that models with relatively 
small $|M_3|$ tend to have good direct detection prospects for this 
reason. In the following, I will establish that this holds in some 
generality in compressed supersymmetry models as parameterized above. 
Ref.~\cite{BBPT} has recently independently found comparable results in a 
couple of somewhat different compressed supersymmetry model lines.
(Ref.~\cite{BBPT} also examines indirect detection opportunities, which 
may
be observable but are subject to much larger uncertainties in
backgrounds and signals.)

The four panels in Figure \ref{fig:DD} have $C_{24} = 0.19$, $0.21$, 
$0.24$, and $0.28$. In each panel, $\tan\beta = 10$ is fixed, and $\mu > 
0$. Then $M_1$ is varied, and $m_0$ is adjusted to the (lower) value that 
gives $\Omega_{\rm DM} h^2 = 0.11$. Model points for which $\tilde N_1 
\tilde N_1 \rightarrow t \bar t$ contributes more than 50\% to 
$1/(\Omega_{\rm DM} h^2)$ are denoted by (blue) X's, and those for which 
$\tilde N_1 \tilde N_1 \rightarrow t \bar t$ contributes less than 50\% 
are denoted by (red) dots. The Higgs mass constraint is imposed, which 
cuts off the model points to the left side and above. 
The resulting smallest allowed values of $-A_0/M_1$ at $M_{\rm GUT}$
are $(0.1, 0.4, 0.7, 1.1)$ for $C_{24} = (0.19, 0.21, 0.24, 0.28)$,
respectively, and correspond to the highest lines of models.
Each lower line of model points corresponds to an increase of $-A_0/M_1$ 
by $0.1$, up to $-A_0/M_1 = 1.6$. Each lower line of models then has
$-A_0/M_1 $ increased by 0.2, up to a limit of $-A_0/M_1 = 
(4.0, 3.8, 3.4, 3.4)$ for $C_{24} = (0.19, 0.21, 0.24, 0.28)$.
These upper limits on $-A_0/M_1$ follow from the requirement of a 
consistent model solution with a stable vacuum and 
$\Omega_{\rm DM} h^2 = 0.11$.

The spin-independent elastic cross-section can clearly be seen to be 
larger for smaller $-A_0$. This is because smaller $-A_0$ corresponds to 
smaller $\mu$, and therefore enhanced bino-higgsino mixing in the LSP. 
Likewise, larger values of $C_{24}$ imply a larger cross-section.

I have checked that repeating the analysis of Figure \ref{fig:DD} but 
with larger values of $\tan\beta$ gives even larger cross-sections, while 
taking smaller values of $\tan\beta$ gives cross-sections that are about 
the same (or only slightly smaller). For the models of most interest 
here, in which $\tilde N_1 
\tilde N_1 \rightarrow t \overline t$ contributes more than 50\% to
$1/(\Omega_{\rm DM} h^2)$, the cross-sections can be up to about 4 times
larger (but often less) than shown in Figure \ref{fig:DD}. However, for fixed values of
the other parameters, increasing $\tan\beta$ within this class of models 
never raises
the cross-section enough to conflict with the present experimental
limit from XENON10.

It is apparent that the next generation of experiments has a good chance 
to detect the dark matter in these models, particularly if $M_1$ and 
$-A_0/M_1$ are not too large, as suggested by subjective naturalness 
criteria. Even taking into account the significant uncertainties in the 
local density of dark matter and the proton matrix elements mentioned 
above, it appears that the ton-scale XENON1T or similar experiments 
should be able to definitively test almost all of the considered models
where $\tilde N_1 \tilde N_1 \rightarrow t \overline t$ makes more than
a 50\% contribution to $1/(\Omega_{\rm DM} h^2)$, 
assuming the local dark matter density and velocity distributions are not 
significantly different from expectations. The exceptions are the models
with the largest possible allowed 
values of $-A_0/M_1 > 3.4$ for $C_{24} = 0.19$ or $0.21$.
However, as observed in the previous 
section, such models can be reasonably considered less likely to be 
realized in nature, since they require relatively larger and more finely 
tuned $\mu$. For the seemingly more natural case 
$-A_0/M_1 \lsim 1.5$, the XENON100 and SuperCDMS 
experiments have a chance to probe most of the parameter space,
particularly for smaller LSP masses.

These results are in contrast to stau 
co-annihilation models, which tend to have a much lower 
spin-independent 
cross-section. Focus point models usually have a comparable or somewhat 
larger cross-section to the ones shown in Figure \ref{fig:DD}. 

\section{Outlook}
\label{sec:outlook}
\setcounter{equation}{0}
\setcounter{footnote}{1}

Compressed supersymmetry, featuring a ratio of the heaviest to lightest 
superpartner masses that is reduced compared to mSUGRA and other models, 
can naturally explain the thermal relic abundance of dark matter in the 
universe while ameliorating the problem of fine-tuning in the 
electroweak symmetry breaking sector of the MSSM. The dark matter 
abundance is related both to the existence of a top squark with mass not 
far above the LSP mass and to a $\mu$ parameter that is not too large 
while maintaining a predominantly bino-like LSP. This scenario has 
distinctive discovery signals at the Large Hadron Collider, as noted in 
\cite{Martin:2007gf} and studied in more detail in \cite{BBPT}, and 
should be discovered despite relatively lepton-poor channels
due to the near
decoupling of sleptons and wino-like charginos from superpartner decay 
chains. Unfortunately, prospects for study at a future $e^+e^-$
linear collider are likely to be severely limited in this scenario 
\cite{Martin:2007gf} unless the center of mass
energy is considerably larger than $\sqrt{s} = 500$ GeV.

In this paper, I have studied the possibility of direct detection of dark 
matter in the top-squark mediated neutralino annihilation case, with 
results consistent with those independently obtained recently in 
\cite{BBPT}. The current limits from CDMS and XENON-10 do not impact on 
the scenario at all. However, I have found that the prospects in a wide 
variety of models of this type look quite promising for the next 
generation of experiments, for models with $-A_0/M_1$ smaller 
than roughly $1.5$, which seem to be the least fine-tuned. 
Better, a 1-ton class XENON experiment (or a 
comparable one) should have the necessary reach to detect 
neutralino scattering events, even taking into account uncertainties in 
the cross-section and the local density of dark matter, except in some 
parts of the model parameter space that would require quite severe 
fine-tuning. One may therefore anticipate that dark matter direct 
detection will be quite useful in confirming and clarifying future LHC 
results if this scenario is correct.


\bigskip \noindent {\it Acknowledgments:} This work was supported in part 
by the National Science Foundation grant number PHY-0456635.



\begin{thebibliography}{90}
\baselineskip=12.3pt

\bibitem{DMreviews}
G.~Jungman, M.~Kamionkowski and K.~Griest,
  Phys.\ Rept.\  {\bf 267}, 195 (1996)
  [hep-ph/9506380],
G.~Bertone, D.~Hooper and J.~Silk,
  Phys.\ Rept.\  {\bf 405}, 279 (2005)
  [hep-ph/0404175].

\bibitem{primer} A review that uses the same notations and conventions
as here is 
S.P.~Martin,
  ``A supersymmetry primer,''
  [hep-ph/9709356], version 4 June 2006.

\bibitem{WMAP}
  D.N.~Spergel {\it et al.}  [WMAP Collaboration],
  ``Wilkinson Microwave Anisotropy Probe (WMAP) three year results:     
  Implications for cosmology,''
  [astro-ph/0603449].

\bibitem{SDSS}
  M.~Tegmark {\it et al.}  [SDSS Collaboration],
  Phys.\ Rev.\  D {\bf 69}, 103501 (2004)
  [astro-ph/0310723].

\bibitem{PDG}
W.M.~Yao {\it et al.}  [Particle Data Group],
  ``Review of particle physics,''
  J.\ Phys.\ G {\bf 33}, 1 (2006).

\bibitem{Drees:1992am}
  M.~Drees and M.M.~Nojiri,
  Phys.\ Rev.\  D {\bf 47}, 376 (1993)
  [hep-ph/9207234].

\bibitem{Edsjo:1997bg}
  J.~Edsj\"o and P.~Gondolo,
  Phys.\ Rev.\  D {\bf 56}, 1879 (1997)
  [hep-ph/9704361].
  
\bibitem{focuspoint}
J.L.~Feng, K.T.~Matchev and F.~Wilczek,
  Phys.\ Lett.\ B {\bf 482}, 388 (2000)
  [hep-ph/0004043],
  K.L.~Chan, U.~Chattopadhyay and P.~Nath,
  Phys.\ Rev.\ D {\bf 58}, 096004 (1998)
  [hep-ph/9710473],
  J.L.~Feng, K.T.~Matchev and T.~Moroi,
  Phys.\ Rev.\ Lett.\  {\bf 84}, 2322 (2000)
  [hep-ph/9908309];
  Phys.\ Rev.\ D {\bf 61}, 075005 (2000)
  [hep-ph/9909334].
  
\bibitem{GriestSeckel}
K.~Griest and D.~Seckel,
  Phys.\ Rev.\ D {\bf 43}, 3191 (1991). 
  
\bibitem{staucoannihilation}
J.R.~Ellis, T.~Falk and K.A.~Olive,
  Phys.\ Lett.\ B {\bf 444}, 367 (1998)
  [hep-ph/9810360];
J.R.~Ellis, T.~Falk, K.A.~Olive and M.~Srednicki,
  Astropart.\ Phys.\  {\bf 13}, 181 (2000)
  [Erratum-ibid.\  {\bf 15}, 413 (2001)]
  [hep-ph/9905481].
  
\bibitem{stopcoannihilationone}
M.E.~Gomez, G.~Lazarides and C.~Pallis,
  Phys.\ Rev.\ D {\bf 61}, 123512 (2000)
  [hep-ph/9907261];
  
\bibitem{stopcoannihilationtwo} 
C.~Boehm, A.~Djouadi and M.~Drees,
  Phys.\ Rev.\ D {\bf 62}, 035012 (2000)
  [hep-ph/9911496];
  
\bibitem{stopcoannihilationthree}
J.R.~Ellis, K.A.~Olive and Y.~Santoso,
  Astropart.\ Phys.\  {\bf 18}, 395 (2003)
  [hep-ph/0112113];

\bibitem{Edsjo:2003us}
J.~Edsjo, M.~Schelke, P.~Ullio and P.~Gondolo,
  JCAP {\bf 0304}, 001 (2003)
  [hep-ph/0301106,
J.~Edsjo, M.~Schelke and P.~Ullio,
  JCAP {\bf 0409}, 004 (2004)
  [astro-ph/0405414].

\bibitem{stopcoannihilationfour}
C.~Balazs, M.~Carena and C.E.M.~Wagner,
  Phys.\ Rev.\ D {\bf 70}, 015007 (2004)
  [hep-ph/0403224].
  
\bibitem{stopcoannihilationfive}
  G.~Belanger, F.~Boudjema, S.~Kraml, A.~Pukhov and A.~Semenov,
  Phys.\ Rev.\  D {\bf 73}, 115007 (2006)
  [hep-ph/0604150].

  
\bibitem{Berezinsky:1995cj} 
  V.~Berezinsky et al.,
  Astropart.\ Phys.\  {\bf 5}, 1 (1996)
  [hep-ph/9508249].

\bibitem{Nath:1997qm}
  P.~Nath and R.~Arnowitt,
  Phys.\ Rev.\  D {\bf 56}, 2820 (1997)
  [hep-ph/9701301].
  
\bibitem{Baer:2001vw}
  H.~Baer, C.~Balazs, M.~Brhlik, P.~Mercadante, X.~Tata and Y.~Wang,
  Phys.\ Rev.\  D {\bf 64}, 015002 (2001)
  [hep-ph/0102156].
  
\bibitem{Ellis:2002iu}
  J.R.~Ellis, T.~Falk, K.A.~Olive and Y.~Santoso,
  Nucl.\ Phys.\  B {\bf 652}, 259 (2003)
  [hep-ph/0210205].
  J.R.~Ellis, A.~Ferstl, K.A.~Olive and Y.~Santoso,
  Phys.\ Rev.\  D {\bf 67}, 123502 (2003)
  [hep-ph/0302032].
  J.R.~Ellis, K.A.~Olive, Y.~Santoso and V.C.~Spanos,
  Phys.\ Lett.\  B {\bf 603}, 51 (2004)
  [hep-ph/0408118].
  
\bibitem{Baer:2004fu}
  H.~Baer, A.~Mustafayev, S.~Profumo, A.~Belyaev and X.~Tata,
  Phys.\ Rev.\  D {\bf 71}, 095008 (2005)
  [hep-ph/0412059].
  
\bibitem{DeRoeck:2005bw}
  A.~De Roeck et al.,
  ``Supersymmetric benchmarks with non-universal scalar masses or gravitino
  dark matter,''
 [hep-ph/0508198].
  
\bibitem{Evans:2006sj}
  J.L.~Evans, D.E.~Morrissey and J.D.~Wells,
  Phys.\ Rev.\  D {\bf 75}, 055017 (2007)
  [hep-ph/0611185].

  
\bibitem{Corsetti:2000yq}
  A.~Corsetti and P.~Nath,
  Phys.\ Rev.\  D {\bf 64}, 125010 (2001)   
  [hep-ph/0003186].
  
\bibitem{winocontentDM}
A.~Birkedal-Hansen and B.D.~Nelson,
  Phys.\ Rev.\ D {\bf 64}, 015008 (2001)
  [hep-ph/0102075],
  Phys.\ Rev.\  D {\bf 67}, 095006 (2003)
  [hep-ph/0211071].
  
\bibitem{Chattopadhyay:2001va}
  U.~Chattopadhyay, A.~Corsetti and P.~Nath,
  Phys.\ Rev.\  D {\bf 66}, 035003 (2002)
  [hep-ph/0201001].
  
\bibitem{Baer:2002by}
  H.~Baer, C.~Balazs, A.~Belyaev, R.~Dermisek, A.~Mafi and A.~Mustafayev,
  JHEP {\bf 0205}, 061 (2002)
  [hep-ph/0204108].
  
\bibitem{Bertin:2002sq}
  V.~Bertin, E.~Nezri and J.~Orloff,
  JHEP {\bf 0302}, 046 (2003)
  [hep-ph/0210034].
  
\bibitem{Chattopadhyay:2003yk}
  U.~Chattopadhyay and D.P.~Roy,
  Phys.\ Rev.\  D {\bf 68}, 033010 (2003)
  [hep-ph/0304108].
  
\bibitem{Cerdeno:2004zj}
  D.G.~Cerde\~no and C.~Mu\~noz,
  JHEP {\bf 0410}, 015 (2004)
  [hep-ph/0405057].
  
\bibitem{Belanger:2004ag}
  G.~Belanger, F.~Boudjema, A.~Cottrant, A.~Pukhov and A.~Semenov,
  Nucl.\ Phys.\  B {\bf 706}, 411 (2005) 
  [hep-ph/0407218].
  
\bibitem{Belanger:2004hk}
  G.~Belanger, F.~Boudjema, A.~Cottrant, A.~Pukhov and A.~Semenov,
  Czech.\ J.\ Phys.\  {\bf 55}, B205 (2005)
  [hep-ph/0412309].
  
\bibitem{Mambrini:2005cp}
  Y.~Mambrini and E.~Nezri,
  Eur.\ Phys.\ J.\  C {\bf 50}, 949 (2007)
  [hep-ph/0507263].
  
\bibitem{King:2006tf}
  S.F.~King and J.P.~Roberts,     
  JHEP {\bf 0609}, 036 (2006)
  [hep-ph/0603095];
  JHEP {\bf 0701}, 024 (2007)  
  [hep-ph/0608135].
  
\bibitem{Baer:2006dz}
  H.~Baer, A.~Mustafayev, E.K.~Park, S.~Profumo and X.~Tata,
  JHEP {\bf 0604}, 041 (2006)
  [hep-ph/0603197],
  H.~Baer, A.~Mustafayev, S.~Profumo and X.~Tata,
  Phys.\ Rev.\  D {\bf 75}, 035004 (2007)
  [hep-ph/0610154].
  
\bibitem{welltempered}
  N.~Arkani-Hamed, A.~Delgado and G.F.~Giudice,
  Nucl.\ Phys.\  B {\bf 741}, 108 (2006)
  [hep-ph/0601041].
  
\bibitem{Falkowski:2005ck}
  A.~Falkowski, O.~Lebedev and Y.~Mambrini,
  JHEP {\bf 0511}, 034 (2005)
  [hep-ph/0507110].
  
\bibitem{Baeretalmirage}
  H.~Baer, E.K.~Park, X.~Tata and T.T.~Wang,
  JHEP {\bf 0608}, 041 (2006)
  [hep-ph/0604253];
  ``Collider and Dark Matter Phenomenology of Models with Mirage Unification,''
  [hep-ph/0703024].
  
\bibitem{Bae:2007pa}
  K.J.~Bae, R.~Dermisek, H.D.~Kim and I.W.~Kim,
  ``Mixed bino-wino-higgsino dark matter in gauge messenger models,''
  [hep-ph/0702041].

\bibitem{King:2007vh}
  S.F.~King, J.P.~Roberts and D.P.~Roy,
  ``Natural Dark Matter in SUSY GUTs with Non-universal Gaugino Masses,''
  [hep-ph/0705.4219].

\bibitem{Martin:2007gf}
  S.P.~Martin,
  ``Compressed supersymmetry and natural neutralino dark matter from top
  squark-mediated annihilation to top quarks,''
  Phys.\ Rev.\  D {\bf 75}, 115005 (2007)
  [hep-ph/0703097].

\bibitem{BBPT} 
  H.~Baer, A.~Box, E.~Park and X. Tata,
  ``Implications of compressed supersymmetry for collider and 
  dark matter searches", 
  [hep-ph/0707.0618]. 
  
\bibitem{KaneKing}
  G.L.~Kane and S.F.~King,
  Phys.\ Lett.\  B {\bf 451}, 113 (1999)
  [hep-ph/9810374],
  M.~Bastero-Gil, G.L.~Kane and S.F.~King,
  Phys.\ Lett.\  B {\bf 474}, 103 (2000)
  [hep-ph/9910506].

\bibitem{TeVtopmass}
The Tevatron Electroweak Working Group for the CDF and D0 collaborations,
  ``A combination of CDF and D0 results on the mass of the top quark,''
  [hep-ex/0703034].

\bibitem{LEPHiggsbounds}
R.~Barate {\it et al.}  [LEP Working Group for Higgs boson searches],
  Phys.\ Lett.\  B {\bf 565}, 61 (2003)
  [hep-ex/0306033].
S.~Schael {\it et al.}  [LEP Working Group for Higgs boson searches],
  Eur.\ Phys.\ J.\  C {\bf 47}, 547 (2006)
  [hep-ex/0602042].


\bibitem{micrOMEGAs} 
  G.~Belanger, F.~Boudjema, A.~Pukhov and A.~Semenov,
  ``micrOMEGAs2.0: A program to calculate the relic density of dark matter in
  a generic model,''
  Comput.\ Phys.\ Commun.\  {\bf 176}, 367 (2007)
  [hep-ph/0607059],
  Comput.\ Phys.\ Commun.\  {\bf 174}, 577 (2006)
  [hep-ph/0405253],
  Comput.\ Phys.\ Commun.\  {\bf 149}, 103 (2002)
  [hep-ph/0112278].

\bibitem{DarkSUSY}
  P.~Gondolo, J.~Edsj\"o, P.~Ullio, L.~Bergstrom, M.~Schelke and E.A.~Baltz,
  ``DarkSUSY: Computing supersymmetric dark matter properties numerically,''
  JCAP {\bf 0407}, 008 (2004)
  [astro-ph/0406204];

\bibitem{softsusy} 
  B.C.~Allanach,
  ``SOFTSUSY: A C++ program for calculating supersymmetric spectra,''
  Comput.\ Phys.\ Commun.\  {\bf 143}, 305 (2002)
  [hep-ph/0104145].

\bibitem{suspect}   
  A.~Djouadi, J.L.~Kneur and G.~Moultaka,
  ``SuSpect: A Fortran code for the supersymmetric and Higgs particle spectrum
  in the MSSM,''
  [hep-ph/0211331].  

\bibitem{ISAJET}
F.E.~Paige, S.D.~Protopopescu, H.~Baer and X.~Tata,
  ``ISAJET 7.69: A Monte Carlo event generator for p p, anti-p p, and e+ e-
  reactions,''
  [hep-ph/0312045].


\bibitem{SMbsgamma} 
K.G.~Chetyrkin, M.~Misiak and M.~Munz,
  Phys.\ Lett.\  B {\bf 400}, 206 (1997)
  [Erratum-ibid.\  B {\bf 425}, 414 (1998)]
  [hep-ph/9612313];
A.J.~Buras, A.~Kwiatkowski and N.~Pott,
  Phys.\ Lett.\  B {\bf 414}, 157 (1997)
  [Erratum-ibid.\  B {\bf 434}, 459 (1998)]
  [hep-ph/9707482];
A.L.~Kagan and M.~Neubert,
  Eur.\ Phys.\ J.\  C {\bf 7}, 5 (1999)
  [hep-ph/9805303].

\bibitem{Barberio:2007cr}
  E.~Barberio {\it et al.}  [Heavy Flavor Averaging Group
                  Collaboration],
  ``Averages of b-hadron properties at the end of 2006,''
  [hep-ex 0704.3575].

\bibitem{Ellis:1985jn}
  J.R.~Ellis, K.~Enqvist, D.V.~Nanopoulos and K.~Tamvakis,
  Phys.\ Lett.\  B {\bf 155}, 381 (1985).
  
\bibitem{Drees:1985bx}
  M.~Drees,
  Phys.\ Lett.\  B {\bf 158}, 409 (1985).
  
\bibitem{Anderson:1996bg}
  G.~Anderson, C.H.~Chen, J.F.~Gunion, J.D.~Lykken, T.~Moroi and Y.~Yamada,
  ``Motivations for and implications of non-universal GUT-scale boundary
  conditions for soft SUSY-breaking parameters,'' report for Snowmass 96,
  [hep-ph/9609457].
  
\bibitem{Anderson:1999ui}
  G.~Anderson, H.~Baer, C.~h.~Chen and X.~Tata,  
  Phys.\ Rev.\  D {\bf 61}, 095005 (2000)
  [hep-ph/9903370].

\bibitem{muonamm}
G.W.~Bennett {\it et al.}  [Muon G-2 Collaboration],
  Phys.\ Rev.\  D {\bf 73}, 072003 (2006)
  [hep-ex/0602035].

\bibitem{IRfp}
I.~Jack and D.R.T.~Jones,
  Phys.\ Lett.\  B {\bf 349}, 294 (1995)
  [hep-ph/9501395].
I.~Jack, D.R.T.~Jones and K.L.~Roberts,
  Nucl.\ Phys.\  B {\bf 455}, 83 (1995)
  [hep-ph/9505242].
M.~Lanzagorta and G.G.~Ross,
  Phys.\ Lett.\  B {\bf 364}, 163 (1995)
  [hep-ph/9507366].

\bibitem{UVfp} 
S.P.~Martin and J.D.~Wells,
  Phys.\ Rev.\  D {\bf 64}, 036010 (2001)
  [hep-ph/0011382].

\bibitem{Atfp}
M.~Carena, M.~Olechowski, S.~Pokorski and C.E.M.~Wagner,
  Nucl.\ Phys.\  B {\bf 419}, 213 (1994)
  [hep-ph/9311222],
M.~Carena and C.E.M.~Wagner,
  Nucl.\ Phys.\  B {\bf 452}, 45 (1995)
  [hep-ph/9408253],
S.A.~Abel and B.~Allanach,
  Phys.\ Lett.\  B {\bf 415}, 371 (1997)
  [hep-ph/9707436].
  

\bibitem{CDMSIIlimit}
D.S.~Akerib {\it et al.}  [CDMS Collaboration],
  Phys.\ Rev.\ Lett.\  {\bf 96}, 011302 (2006)
  [astro-ph/0509259].

\bibitem{XENONten}
J.~Angle {\it et al.}  [XENON Collaboration],
  ``First Results from the XENON10 Dark Matter Experiment at the Gran Sasso
  National Laboratory,''
  [astro-ph/0706.0039].

\bibitem{XENON} 
E.~Aprile {\it et al.},
  ``XENON: A 1-tonne liquid xenon experiment for a sensitive dark matter
  search,''
  [astro-ph/0207670];
  Nucl.\ Phys.\ Proc.\ Suppl.\  {\bf 138}, 156 (2005)
  [astro-ph/0407575].

\bibitem{SuperCDMS} 
D.S.~Akerib {\it et al.},
  ``The SuperCDMS proposal for dark matter detection,''
  Nucl.\ Instrum.\ Meth.\  A {\bf 559}, 411 (2006).

\bibitem{DMtools} Direct dark matter detection experiment limits and 
projections may be found at\hfill\\
{\tt http://dendera.berkeley.edu/plotter/} and
in the directory {\tt plotter/limitdata}.

\bibitem{Drees:1993bu}
  M.~Drees and M.~Nojiri,
  ``Neutralino - Nucleon Scattering Revisited,''
  Phys.\ Rev.\  D {\bf 48}, 3483 (1993)
  [hep-ph/9307208].

\bibitem{DDuncertainties}
A.~Bottino, F.~Donato, N.~Fornengo and S.~Scopel,
  Astropart.\ Phys.\  {\bf 13}, 215 (2000)
  [hep-ph/9909228].
J.R.~Ellis, K.A.~Olive, Y.~Santoso and V.C.~Spanos,
  Phys.\ Rev.\  D {\bf 71}, 095007 (2005)
  [hep-ph/0502001].

\bibitem{matrixelements}
J.~Gasser, H.~Leutwyler and M.~E.~Sainio,
  Phys.\ Lett.\  B {\bf 253}, 252 (1991).
  
\end{thebibliography}
\end{document}